\documentclass[a4paper]{amsart}
\usepackage{amsmath}
\usepackage{amsthm}
\usepackage{amssymb}
\usepackage[mathscr]{eucal} 
\usepackage{amscd}
\usepackage{epsfig}

\newtheorem{Thm}{Theorem}[section]
\newtheorem{Cor}[Thm]{Corollary}
\newtheorem{Prop}[Thm]{Proposition}
\newtheorem{Lemma}[Thm]{Lemma}
\theoremstyle{definition}

\theoremstyle{remark}
\newtheorem{rem}[Thm]{Remark}

\numberwithin{equation}{section}
\newcommand{\calA}{{\mathcal A}}

\newcommand{\calB}{{\mathcal B}}

\newcommand{\calC}{{\mathcal C}}
\newcommand{\calD}{{\mathcal D}}

\newcommand{\De}{\Delta}

\newcommand{\calG}{{\mathcal G}}
\newcommand{\calP}{{\mathcal P}}

\newcommand{\f}{\varphi}

\newcommand{\calH}{{\mathcal H}}
\newcommand{\frH}{{\mathfrak H}}
\newcommand{\calI}{{\mathcal I}}
\newcommand{\calK}{{\mathcal K}}

\newcommand{\calL}{{\mathcal L}}
\newcommand{\calM}{{\mathcal M}}
\newcommand{\m}{\mu}
\newcommand{\calN}{{\mathcal N}}

\newcommand{\calO}{{\mathcal O}}

\newcommand{\Q}{\Omega}
\newcommand{\g}{\gamma}

\newcommand{\R}{{\mathbb R}}

\newcommand{\calU}{{\mathcal U}}
\newcommand{\calV}{{\mathcal V}}
\newcommand{\calW}{{\mathcal W}}

\newcommand{\calZ}{{\mathcal Z}}
\newcommand{\Ze}{{\mathbb Z}}

\DeclareMathOperator*{\Ad}{Ad}
\DeclareMathOperator*{\Iso}{Iso}
\def\eps{\varepsilon}

\def\L{\Lambda}

\def\a{\alpha}
\def\b{\beta}
\def\man{\calM}
\def\CONF{\mathrm{CONF}}
\def\Conf{\mathrm{Conf}}
\def\conf{\mathrm{conf}}
\def\Iso{\mathrm{Iso}}

\begin{document}
\author[D. Guido, R. Longo]{{\bf 
Daniele Guido and Roberto Longo}\\
\hfill\\
Dipartimento di Matematica,
Universit\`a di Roma ``Tor Vergata''\\
Via della Ricerca Scientifica 1, I-00133 Roma, Italy}

\title
[Converse Hawking-Unruh Effect]
{$\text{\huge  A Converse Hawking-Unruh Effect}$ \\
\vskip0.1cm
$\text{\huge and $\mathbf{dS^2/CFT}$ Correspondence}$} 

\date{}

\thanks{e-mails: guido@mat.uniroma2.it,
longo@mat.uniroma2.it} 

\thanks{Supported in part by MIUR and
GNAMPA-INDAM} 

\thanks{This paper has been circulated under the title ``The Cooling
of Hawking Radiation and dS/CFT Correspondence''}
\begin{abstract} 
    Given a local quantum field theory net $\calA$ on the de Sitter
    spacetime $dS^d$, where geodesic observers are thermalized at
    Gibbons-Hawking temperature, we look for observers that feel to be
    in a ground state, i.e. particle evolutions with positive
    generator, providing a sort of converse to the Hawking-Unruh
    effect.  Such positive energy evolutions always exist as
    noncommutative flows, but have only a partial geometric meaning,
    yet they map localized observables into localized observables.

    We characterize the local conformal nets on $dS^d$.  Only in this
    case our positive energy evolutions have a complete geometrical
    meaning.  We show that each net has a unique maximal expected
    conformal subnet, where our evolutions are thus geometrical.

    In the two-dimensional case, we construct a holographic one-to-one
    correspondence between local nets $\calA$ on $dS^2$ and local
    conformal non-isotonic families (pseudonets) $\calB$ on $S^1$. 
    The pseudonet $\calB$ gives rise to two local conformal nets
    $\calB_{\pm}$ on $S^1$, that correspond to the $\frH_{\pm}$ horizon
    components of $\calA$, and to the chiral components
    of the maximal conformal subnet of $\calA$.  In particular,
    $\calA$ is holographically reconstructed by a single horizon
    component, namely the pseudonet is a net, iff the translations on
    $\mathfrak H_{\pm}$ have positive energy and the translations on
    $\mathfrak H_{\mp}$ are trivial.  This is the case iff the
    one-parameter unitary group implementing rotations on $dS^{2}$ has
    positive/negative generator.
\end{abstract}
\maketitle
{\footnotesize \tableofcontents}

\section{Introduction}
\label{sec:intro}
The thermalization effects discovered by Hawking \cite{Haw}, resp. by 
Unruh \cite{U}, have shown the concept of particle to be gravity, 
resp. observer, dependent; the two effects being related by Einstein 
equivalence principle. 

Unruh effect deals in particular with a quantum field theory on
Minkowski spacetime: an observer $O$ with uniform acceleration $a$
feels, in its proper Rindler spacetime $W$, the Hawking temperature
$T_{\text{H}}=\frac{a}{2\pi}$.  As noticed in \cite{S}, this can be
also explained by the Bisognano-Wichmann theorem \cite{BiWi2}: the
one-parameter automorphism group describing the evolution of $O$ in
its proper observable algebra $\calA(W)$ satisfies the KMS thermal
equilibrium condition at inverse temperature $\beta _{\text{H}}
=T^{-1}_{\text{H}}$, see \cite{Haa}.

The Gibbons-Hawking effect \cite{GH} occurs in the de Sitter spacetime
with radius $\rho$.  Here every inertial observer $O$ feels the
temperature $T_{\text{GH}}=\frac{1}{2\pi\rho}$.  Again we may express
this fact by saying that the one-parameter automorphism group
describing the evolution of $O$ in its proper observable algebra
$\calA(W)$ satisfies the KMS thermal equilibrium condition at inverse
temperature $\beta_{\text{GH}} =T^{-1}_{\text{GH}}$, where $W$ is here
the static de Sitter spacetime (\cite{FHN} for the two-dimensional
case).

In the Minkowski spacetime, the KMS property for a uniformly
accelerated observer $O$ can be taken as a first principle, then the
basic structure follows, in particular the Poincar\'e symmetries with
positive energy are then derived \cite{GuLo1,BGL2,gl:95,GL4} and
\cite{BS,BDFS} for a related approach. See also
\cite{GL4,Ku3} for weaker thermal conditions.

In the de Sitter space, the KMS property for the geodesic observer $O$
can be taken as a first principle \cite{BoBu,BEM,NPT}; in particular
the value of $T_{\text{GH}}$ is then fixed automatically \cite{BoBu}.

We mention at this point that actual observations in cosmology
indicates that, on a large scale, our universe is isotropic,
homogeneous and repulsively expanding.  The de Sitter spacetime thus
provides a good approximation model, at least asymptotically.  Hence
de Sitter spacetime and, more generally, Robertson-Walker spacetimes
with positive scalar curvature are basic objects to be be studied.

The first aim of this paper is to study a sort of converse to the
above mentioned thermalization effects.  Starting with the curved de
Sitter space, where a geodesic observer is thermalized, we wish to
find a different observer whose quantum evolution has positive
generator, namely feels to be in a ground state.  In other words, we
want to keep the same state, but choose a time evolution w.r.t. which
the state becomes a ground state.  Now an observer in $dS^d$ whose
world line is an orbit of a boost experiences a temperature
\[
T= \frac{1}{2\pi}\sqrt{\frac{1}{\rho^2} + a^2} \geq T_{\text{GH}},
\]
\cite{NPT}, with $a$ the modulus of the intrinsic uniform
acceleration, contrary to our aim.

The dethermalization effect to take place is indeed a non-trivial
matter.  To understand this point notice that we are looking for a
particle whose acceleration compensates the curvature of the
underlying space so that, at least locally, the particle's picture of
the spacetime is flat.  Yet the particle's acceleration is a vector,
but the curvature is a tensor so that, even in the constant scalar
curvature case, there is no obvious way to fulfil the above
requirement.

Indeed, it turns out that this cooling down effect is linked to the
conformal invariance and, in two spacetime dimensions, to a holography
in a sense similar to the one studied in the anti-de Sitter spacetime
\cite{MaWi}, as we shall explain.

As is known, Minkowski spacetime $M^d$ is conformal to a double cone
in the Einstein static universe $E^d$.  On the other hand $dS^d$ is
conformal to a rectangular strip of $E^d$.  Using this fact one can
directly set up up a bijective correspondence between local conformal
nets on $M^d$ and on $dS^d$.  Less obviously, this sets up a
correspondence between positive energy-momentum local conformal nets
on $M^d$ and local conformal nets on $dS^d$ with the KMS property for
geodesic observers.

At this point it is immediate that, given any local, conformal, KMS
geodesic net on $dS^d$, there exists a timelike conformal geodesic
flow $\mu$ on $dS^d$ that gives rise to a quantum evolution with
positive generators: they are simply the ones that correspond to
timelike translations on $M^d$.  Let us remark that $\m$ has only a
local action on $dS^{d}$, namely in general $\m_{t}x$ ``goes outside''
$dS^{d}$ for large $t$.

We may ask whether the flow $\mu$ promotes to a ground state quantum
evolution for a general local net on $dS^d$.  In a sense we need to
proceed similarly to Mechanics when one passes from a passive
description (in terms of coordinates) to an active description (in
terms of tensors).  The answer is yes, but the evolution is only
partially geometric.  We shall show that there exists a one-parameter
unitary group $V$ with positive generator such that, in particular,
\[
V(t)\calA(\calO)V(-t) = \calA(\mu_t\calO)
\]
for certain regions $\calO$ and for all $t\in\mathbb R$ such that 
$\m_{t}\calO$ is still in $dS^{d}$, and
\[
V(t)\calA(\calO)V(-t)\subset \calA(\tilde\calO'_t),
\]
for all double cones $\calO$ contained in the steady state universe
subregion $\calN$ of $dS$ (or contained in the complement of $\calN$),
for a suitable double cone $\tilde\calO_t$ depending on $\calO$ and
$t$, cf.  Remark \ref{localization}.  The unitary group $V$ is
constructed by the Borchers-Wiesbrock methods \cite{Bo1,Bo2,W} and, in
the conformal case, coincides with the previously considered one where
the geometric meaning is complete.

Our analysis then proceeds to determine the maximal subnet of $\calA$
where the geometric meaning is complete.  For any net $\calA$ we show
that there exists a unique maximal expected conformal subnet, and 
this net has the required property.

Finally we consider more specifically the case of a two-dimensional de
Sitter spacetime, with the aim of describing a local net on $dS^{2}$
via holographic reconstruction, namely in terms of a suitable
conformal theory on $S^{1}$.  For different approaches to $dS/CFT$ 
correspondence in the two-dimensional or in the higher dimensional case, 
see e.g. \cite{St}.

To this end we introduce the notion of pseudonet on $S^1$.  This is a
family of local von Neumann algebras associated with intervals of
$S^1$ where isotony is \emph{not} assumed.  Moreover, we assume
M\"obius covariance, commutativity between the algebra of an interval
and that of its complement, the existence of an invariant cyclic
(vacuum) vector, and the geometric meaning of the modular groups.

We shall show a that a local conformal pseudonet $\calB$ encodes
exactly the same information of a $SO_0(2,1)$-covariant local net
$\calA$ on $dS^2$ with the geodesic KMS property, namely we have a
bijective correspondence, holography,
\[
SO_0(2,1)-\text{\emph{covariant local nets on $dS^2$}}\leftrightarrow
\text{\emph{local conformal pseudonets on $S^1$}.}
\]

The pseudonet naturally lives on one component $\frH_{+}$ or
$\frH_{-}$ of the cosmological horizon (choosing the other horizon
component would amount to pass to the conjugate pseudonet), and the
holographic reconstruction is based on a $1:1$ geometric
correspondence between wedges in $dS^{2}$ and their projections on
$\frH_{\pm}$.  Conformal invariance and chirality may then be
described in terms of the pseudonet.

A net $\calA$ on $dS^2$ gives, by restriction, two nets $\calA_{\pm}$
on $\mathfrak H_{\pm}$, that turn out to be conformal, hence
$\calA_{\pm}$ extend to conformal nets on $S^1$.  Then $\calA_{\chi}:=
\calA_{+}\otimes\calA_{-}$ is a two-dimensional chiral conformal net. 
It turns out that $\calA_{\chi}$ is naturally identified with a
conformal subnet of $\calA$, indeed it is the chiral subnet of the
maximal conformal subnet of $\calA$.

From a different point of view, the pseudonet $\calB$ gives naturally
rise to a pair of local conformal (i.e. M\"obius covariant) nets
$\calB_{\pm}$ on $S^1$ that correspond to $\calA_{\pm}$.

Finally we address the question of when a net on $dS^{2}$ is
holographically reconstructed by a conformal net, namely when the
associated pseudonet is indeed isotonic.  Let $\tau$ be the Killing
flow (\ref{param1}) which restricts to the translations on $\mathfrak
H_+$, and $U$ is the associated one-parameter unitary subgroup of the
de Sitter group representation.  Then $U$ has positive generator if
and only if $\calB$ is isotonic.  This is perhaps a point where the
relation between the dethermalization effect, conformal invariance and
holography is more manifest.  Indeed the two-dimensional case is the
only case where the de Sitter group admits positive energy
representations.  This means that a massless particle on $\frH_{+}$
may evolve according with the flow $\tau$ (cf.  subsection \ref{tau})
hence feels a dethermalized vacuum if the representation is positive
energy.  However this is exactly the case where the net is conformal
and ``lives on $\frH_{+}$'', namely the restriction $\calA_-$ is
trivial.  Analogous result holds replacing $\mathfrak H_+$ with
$\mathfrak H_-$.

\section{General Structure} 
\label{sec2}

\subsection{Geometrical preliminaries} We begin to recall some basic 
structure, mainly geometrical aspects, that will undergo our analysis.

\subsubsection{Expanding universes and Gibbons-Hawking effect} As is
known \cite{GH}, a spacetime $\calM$ with repulsive (i.e. positive)
cosmological constant has certain similarities with a black hole
spacetime.  $\calM$ is expanding so rapidly that, if $\gamma$ is a
freely falling observer in $\calM$, there are regions of $\calM$ that
are inaccessible to $\gamma$, even he waits indefinitely long; in
other words the past of the world line of $\gamma$ is a proper
subregion $\calN$ of $\calM$.  The boundary $\mathfrak H$ of $\calN$
is a cosmological event horizon for $\gamma$.  As in the black hole
case, one argues that $\gamma$ detects a temperature related to the
surface gravity of $\mathfrak H$.  This is a quantum effect described
by quantum fields on $\calM$ (see below); heuristically: spontaneous
particle pairs creation happens on $\mathfrak H$, negative energy
particles may tunnel into the inaccessible region, the others
contribute to the thermal radiations.

\subsubsection{de Sitter spacetime} The spherically symmetric,
complete vacuum solution of Einstein equation with cosmological
constant $\Lambda > 0$ is $dS^d$, the $d$-dimensional de Sitter
spacetime.  By Hopf theorem, if $d>2$, $dS^d$ is the unique complete
simply connected spacetime with constant curvature $R=2d\Lambda/(d-2)$ (if
$d=2$ this characterizes the universal covering of $dS^2$).  $dS^d$
may be defined as a pseudosphere, namely the submanifold of the
ambient Minkowski spacetime $\mathbb R^{d+1}$
\[
x_0^2 - x_1^2 - \cdots - x_d^2 =  - \rho^2
\]
where the de Sitter radius is
$\rho=\sqrt{\frac{(d-1)(d-2)}{2\Lambda}}$.  $dS^d$ is maximally
symmetric, isotropic and homogeneous; the de Sitter group $SO(d,1)$
acts transitively by isometries of $dS^d$.  The geodesics of $dS^d$
are obtained by intersecting $dS^d$ with two-dimensional planes
through the origin of $\mathbb R^{d+1}$, see e.g. \cite{Ri,ON}.  In
particular the world line of a material freely falling observer is a
boost flow line, say
\begin{equation}\label{geo}
\left\{ \aligned
x_0(t) &=  \rho\sinh t \\
x_1(t) &=  \rho\cosh t\\
x_2(t) &= 0 \\
\cdots \\
x_d(t) &= 0
\endaligned \right.
\end{equation}
whose past is the steady state universe, the
part of $dS^d$ lying in the region $\calN = \{x: x_1 >
x_0\}$ and the cosmological horizon $\mathfrak H$ is the intersection
of $dS^d$ with the plane $\{x: x_0=x_1\}$.

The orbit of uniformly accelerated observers are obtained by
intersecting $dS^d$ with arbitrary planes of $\mathbb R^{d+1}$
\cite{Ri2}, of course only timelike and lightlike sections describe
material and light particles, the others have constant imaginary
acceleration.

\subsubsection{Killing flows}  
We briefly recall a few facts about the proper spacetime and the 
corresponding evolution of an observer.
Let $\calM$ be a Lorentzian manifold and
$\g:\mathbb R\to\calM$ a timelike or lightlike geodesic. The proper 
spacetime of the observer associated with $\g$ is the causal 
completion $W$ of $\g$.  The
relative acceleration of nearby particles is measured by the second
derivative of the variation vector field $V$ on $\g$; by definition,
if $x: \mathbb R\times (-\delta,\delta)\to\calM$ is a smooth map with
$\g(u)=x(u,0)$ then $V(u)\equiv \partial_v x(u,v)|_{v=0}$.  If $x$ is
geodesic, namely every map $u\to x(u,v)$ is a geodesic, then $V$ is a
Jacobi vector field, namely $V''=R_{V\g'}\g'$ where $R$ is the
curvature tensor, showing that in general there is a non-zero tidal
force $R_{V\g'}\g'$ (we use proper time parametrization in the
timelike case).

On the other hand, if all maps $u\to x(u,v)$ are flow lines of a Killing 
flow $\tau$, and $x(u,v)=\tau_u(x(0,v))$, then the tidal forces vanish. 
Indeed $V(u)$ is the image of $V(0)$ under the differential of 
$\tau_u$, thus $V(u)$ parallel to $V(0)$ because $\g$ is geodesic. 
Therefore the relative velocity, hence the relative acceleration, is $0$.

In other words a Killing flow having the geodesic $\g$ as a flow line 
describes an evolution which is static with respect to the freely 
falling particle associated with $\g$. We shall consider, in 
particular, the case where $\calM$ is $dS^d$ and $\g$ is a boost line; 
then $W$ is, by definition, a \emph{wedge} and the evolution associated with $\g$ is
described by the same one-parameter subgroup of the de Sitter group.

\subsubsection{The two Killing flows of a lightlike particle}
\label{tau}
We consider now a null geodesic $\g$ in $dS^d$. It 
lies in a section by a two-dimensional plane of $\mathbb R^{d+1}$ 
through the 
origin that contains a lightlike vector of $dS^d$. By the 
transitivity of the $SO_0(d,1)$-action, we may assume the plane 
is $\{x: x_0=x_1, x_2 = \cdots = x_n =0\}$.

Contrary to the timelike geodesic situation (\ref{geo}), which is the
flow of a unique Killing flow (boosts), there are here two possible
Killing flows with an orbit in this section.  As $\g$ is lightlike,
there is no proper time associated with $\g$.  We may parametrize
$\g$, for example, as $\g_1(s)=x(s)$ with
\begin{equation}\label{param1}
\left\{ \aligned
x_0(s) &=  s \\
x_1(s) &=  s\\
x_2(s) &= 0 \\
\cdots \\
x_d(s) &= 0, \quad s\in\mathbb R,
\endaligned \right.
\end{equation}
or $\g_2(t)=x(t)$ with
\begin{equation}\label{param2}
\left\{ \aligned
x_0(t) &=  e^t \\
x_1(t) &=  e^t\\
x_2(t) &= 0 \\
\cdots \\
x_d(t) &= 0, \quad t\in\mathbb R ,
\endaligned \right.
\end{equation}
namely $\g_2(t)=\g_1(e^t)$.  The supports of the two curves are of
course different, one is properly contained in the other: 
in the first case it is the entire line, while in
the second case it is only a half-line.

Now $s\to\g_2(s)$ is a flow line of the boosts (\ref{geo}), the 
observable algebra is $\calA(W)$ with $W$ the wedge as above, and we 
are in the Hawking-Unruh situation. The boundary of $W$ is a ``black 
hole'' horizon for the boosts: the observer associated with $\g_2$
cannot send a signal out of $W$ and get it back.

Also $t\to\g_1(t)$ is the flow line of a Killing flow $\tau$. If 
$d=2$ we may use the usual identification of a point 
$(x_0,x_1,x_2)\in\mathbb R^3$
with the matrix 
$\tilde x=\bigl(\begin{smallmatrix} x_0 + x_2& x_1\\
x_1 & x_0 - x_2\end{smallmatrix}\bigr)$, 
so that 
the determinant of $\tilde x$ is the square of the Lorentz length 
of $x$. Now $PSL(2,\mathbb R)$ acts on $\mathbb R^3$ by 
the adjoint map $A\in PSL(2,\mathbb R)\mapsto \Ad A \in SO(2,1)$ where
\[
\Ad A: \tilde x\to 
A\tilde x A^{\text{T}}.
\]
The map $\Ad$ is an isomorphism of $PSL(2,\mathbb R)$ with
$SO_0(2,1)$, the connected component of the identity of $SO(2,1)$, (we
often identify $PSL(2,\mathbb R)$ and $SO_0(2,1)$) and the flow $\tau$
is given by
\begin{equation}\label{flow}
\tau_t =\Ad\begin{pmatrix} 1 & t\\
0 & 1\end{pmatrix}: \tilde x\to 
\begin{pmatrix} 1 & t\\
0 & 1\end{pmatrix}\tilde x \begin{pmatrix} 1 & 0\\
t & 1\end{pmatrix}.
\end{equation}
(If $d>2$, $\tau$ is given by the same formula, but acts trivially on 
the $x_k$ coordinate, $k>2$.)

\begin{Prop} The above flow $\tau$ is the unique Killing flow having 
the curve (\ref{param1}) as a flow line. 
$\tau$ is lightlike on $\mathfrak H$ and otherwise spacelike.
\end{Prop}
\begin{proof}
The statement is proved by elementary computations.
\end{proof}
In this case the observable algebra of $\g_2$ is 
$\calA(\calN)=\calA(dS^d)$ ($=B(\calH)$ in the irreducible case), 
and $\tau$ acts on the cosmological 
event horizon $\mathfrak H=\{x:x_1=x_0\}$, 
the boundary of $\calN=\{x:x_1>x_0\}$.

\subsection{Quantum fields and local algebras}
\label{QFTondS} 
So far we have mainly discussed geometrical aspects of $dS^d$.  We now
consider a quantum field on $dS^d$, but we assume that back reactions
are negligible, namely the geometry of $dS^d$ is not affected by the
field.  

Let us denote by $\calK$ the set of \emph{double cones} of $dS^d$,
namely $\calK$ is the set of non-empty open regions of $dS^d$ with
compact closure that are the intersection of the future of $x$ and the
past of $y$, where $x,y\in dS^d$ and $y$ belongs to the future of $x$. 
A wedge is the limit case where $x$ and $y$ go to infinity.  We shall
denote by $\calW$ the set of wedges and by $\tilde\calK$ the set of
double cones, possibly with one or two vertex at infinity, thus
$\tilde\calK\supset\calK\cup\calW$.  Elements of $\tilde\calK$ are
obtained by intersecting a family of wedges.

The field is described by a (local) net $\calA$ with the following
properties. 
\smallskip

\noindent {\bf a) Isotony and locality.} $\calA$ is an \emph{inclusion
preserving} map
\begin{equation}
\calO \in\calK\rightarrow \calA(\calO)
\end{equation}
 from double cones $\calO \subset dS^d$ to von Neumann algebras
 $\calA(\calO)$ on a fixed Hilbert space $\calH$. 
 $\calA(\calO)$ is to be interpreted as the algebra generated 
 by all observables which can
 be measured in $\calO$.  
 
  For a more general region $D \subset dS^d$
 the algebra $\calA(D)$ is defined as the von Neumann algebra
 generated by the local algebras $\calA(\calO)$ with $\calO \subset 
 D$, $\calO\in\calK$. 
 
The local algebras are supposed to satisfy the condition of
 \emph{locality}, i.e.
\begin{equation}
\calA ( \calO_1 ) \subset \calA ( \calO_2)' \ \ \mbox{if} \ \ \calO_1 \subset \calO_2',
\end{equation} 
where $\calO'$ denotes the spacelike complement of $\calO$ in $dS$ and 
$\calA(\calO)'$ the commutant of $\calA(\calO)$ in $\calB (\calH) $. 
\smallskip

\noindent {\bf b) Covariance.} There is a continuous unitary
representation $U$ of the de Sitter group $SO_0(d,1)$ on $\calH$ such
that for each region $\calO \subset dS^d$
\begin{equation}
U(g)\calA(\calO)U(g)^{-1} = \calA (g\calO), \quad g\in SO_0(d,1). 
\end{equation} 
\smallskip 
\noindent {\bf c) Vacuum with geodesic KMS property.} There is a unit
vector $\Omega \in \calH$, the vacuum vector, which is $U$-invariant
and cyclic for the global algebra $\calA(dS^d)$.  The corresponding
vector state $\omega$ given by
\begin{equation}
\omega(A) = (\Omega, A\Omega), 
\end{equation} 
has the following geodesic KMS-property (see \cite{BoBu}): For every
wedge $W$ the restriction $\omega\!\upharpoonright_{\calA(W)}$
satisfies the KMS-condition at some inverse temperature $\beta > 0$
with respect to the time evolution (boosts) $\Lambda_{W}(t), \, t \in
\mathbb R$, associated with $W$.  In other words, for any pair of
operators $A,B \in \calA(W)$ there exists an analytic function $F$ in
the strip $D\equiv\{ z \in \mathbb C : 0 < \mbox{Im} z < \beta \}$,
bounded and continuous on the closure $\bar D$ of $D$, such that
\begin{equation}
F(t) = \omega (A \alpha_t(B) ), \ \ 
F(t + i\beta) = \omega (\alpha_t (B) A), \quad t \in \mathbb R,
\end{equation}
where $\alpha_t = \Ad U(\Lambda_{W}(t))$.
\smallskip

\noindent {\bf d) Weak additivity.} For each open region $\calO
\subset dS$ we have
\begin{equation}
\bigvee_{g \in SO_0(d,1)} \!\!
\calA (g\calO)  = \calA(dS^d) \ ,
\end{equation}
where the lattice symbol $\vee$ denotes the generated von Neumann
algebra.
\begin{Prop} \cite[Borchers-Buchholz]{BoBu}
\label{BB}
The following hold:
\begin{itemize} 
\item \emph{Reeh-Schlieder property}: $\Omega$ is cyclic for
$\calA(\calO)$ for each fixed open non-empty region $\calO$ (hence it
is separating for $\calA(\calO)$ if the interior of $\calO'$ is
non-empty).

\item \emph{Wedge duality}: For each wedge $W$ we have $\calA(W)'=\calA(W')$.

\item \emph{Gibbons-Hawking temperature}: The inverse temperature is 
$\beta=2\pi\rho$. 

\item \emph{PCT symmetry}: The representation $U$ of $SO_0(d,1)$ 
extends to a (anti-)\- unitary representation of $SO(d,1)$ acting 
covariantly on $\calA$.
\end{itemize}
\end{Prop}
The Reeh-Schlieder property is obtained by using the KMS property 
in place of the analyticity due to the positivity of the energy in the 
usual argument in the Minkowski space. 

Wedge duality then follows as usual by the geometric action of the modular 
group due to Takesaki theorem; that is to say, if $D=W$ is a wedge
and $\calL\subset\calA(W)$ is a von Neumann 
algebra cyclic on $\Omega$ and globally stable under Ad$\L_W$, then 
$\calL=\calA(W)$; this is a know fact in Minkowski spacetime, see e.g. 
\cite{BGL2}.
Note that this argument also shows that the definition of the von 
Neumann algebra $\calA(W)$ is univocal if $W$ is a wedge.

Concerning the construction of the PCT anti-unitary, a corresponding 
result in the Minkowski space is contained in \cite{gl:95}.
\begin{Lemma}\label{wa}
Let $\calA$ satisfy {\bf a)}, {\bf b)} and {\bf c)}. Then
$\calA$ is weakly additive iff the Reeh-Schlieder property holds.
\end{Lemma}
\begin{proof}
Because of Prop.  \ref{BB} it is sufficient to show that if
$\calO\in\calK$ and $\overline{\calA(\calO)\Q}=\calH$ then the von
Neumann algebra $\calL$ generated by the union of $\calA(g\calO)$,
$g\in SO_0(d,1)$, is equal to $\calA(dS^d)$.

Now $\calL\supset \vee_t\calA(\Lambda_W(t)\calO)$ and the latter is 
equal $\calA(W)$ by Takesaki theorem. Thus $\calL\supset \calA(W_1)$ 
for every wedge $W_1$ because $SO_0(d,1)$ acts transitively on $\calW$ 
and we conclude $\calL=\calA(dS^d)$ because every double cone is 
contained in a wedge.
\end{proof}
It follows as in \cite[Prop.  3.1]{gl:95} (see also \cite{BoBu}) that
the center $\calZ$ of $\calA(W)$ coincides with the center of
$\calA(dS^d)$ and $\calA$ has a canonical disintegration, along
$\calZ$, into (almost everywhere) irreducible nets.  Moreover $\calA$
is \emph{irreducible}, i.e. $\calA(dS^d)=B(\calH)$, if and only if
$\Omega$ is the \emph{unique $U$-invariant vector} (see also
\cite{GL3}).

We shall say that $\calA$ satisfies \emph{Haag duality} if
\[
\calA(\calO)'=\calA(\calO')
\]
for all double cones $\calO\in\calK$.

Now it is elementary to check that every double cone is an 
intersection of wedges, indeed
\[
\calO = \bigcap \calW_{\calO}, \quad \calO\in\tilde\calK,
\]
where $\calW_{\calO}$ denotes the set of wedges containing $\calO$.

We then define the \emph{dual net} $\hat\calA$ as
\[
\hat\calA(\calO)\equiv \bigcap_{W\in\calW_{\calO}}\!\calA(W)\ ,
\]
Note that $\hat\calA(W)=\calA(W)$ if $W$ is a wedge, hence
$\hat\calA(D)=\calA(D)$ if every 
double cone $\calO\subset D$ is contained in a wedge $W\subset D$ 
(this is the case if $D$ is union of wedges).

By wedge duality the  net $\hat\calA$ is local (two spacelike 
separated double cones are contained in two spacelike separated 
wedges) and satisfies all properties {\bf a)}-{\bf d)}.

The following proposition is the version of a known fact in Minkowski 
space, cf. \cite{R}. 

\begin{Prop}\label{Haag duality}
$\hat\calA$ is Haag dual:
\[
\hat\calA(\calO)'=\hat\calA(\calO') \ (=\calA(\calO')),\  
\calO\in\calK\ ,
\]
and $\calA=\hat\calA$ iff $\calA$ satisfies Haag duality.
\end{Prop}
\begin{proof}
Let $\{W_i\}$ be the set of wedges in $\calW_{\calO}$.  Then
$\hat\calA(\calO)=\cap_i\hat\calA(W_i)$, hence $\hat\calA(\calO)' =
(\cap_i\calA(W_i))' = \vee_i\calA(W'_i) \subset \hat\calA(\calO')
\subset \hat\calA(\calO)'$.

To check the last part, it sufficient to assume that $\calA$ satisfies
Haag duality and show that $\calA=\hat\calA$.  Indeed in this case
$\calA(\calO)=\calA(\calO')'=(\vee_i\calA(W'_i))'=\cap_i\calA(W_i)=
\hat\calA(\calO)$.
\end{proof}
If $\tau$ is a flow in $dS^d$, in general we may expect a 
corresponding quantum evolution only if $\tau$ static,
namely if $\tau$ is a Killing flow (see Subsect. \ref{tau}).
In this case there is a
one-parameter unitary group $U$ on $\mathfrak H$ implementing $\tau$ 
covariantly:
\[
U(t)\calA(\calO)U(t)^*=\calA(\tau_t\calO).
\]
Indeed $U$ is a one-parameter subgroup of the unitary representation 
of $SO_0(d,1)$, the connected component of the identity in $SO(d,1)$.

In particular, if $\g:u\in\mathbb R\mapsto \g(u)\in dS^d$ is a
timelike or lightlike geodesic, the evolution for the observer
associated with $\g$ is given by a Killing flow having $\g$ in one
orbit.  Now the observable algebra associated with $\g$ is $\calA(W)$,
where $W$ is the causal envelope of $\g$, which is globally invariant
with respect to $\tau$.  If $\g$ describes a material particle, namely
$\tau$ is a boost, then $W$ is the wedge region globally invariant
with respect to such boosts.  By the geodesic KMS property, Ad$U$ is a
one-parameter automorphism group of $\calA(W)$ that satisfies the KMS
thermal equilibrium condition at temperature $\frac{1}{2\pi\rho}$
\cite{FHN,BEM} and this corresponds, as is known, to the Hawking-Unruh
effect \cite{Haw,U}.

\subsubsection{Subnets}
Given a net $\calA$ on $dS^d$ on a Hilbert space $\calH$, 
namely $\calA$ satisfies properties {\bf a)}, {\bf b)}, 
{\bf c)}, {\bf d)}, we shall say the $\calB$ is a \emph{subnet} of 
$\calA$ if 
\[
\calB:\calO\in\calK\to\calB(\calO)\subset\calA(\calO)
\]
is a isotonic map from double cones to von Neumann algebras such 
that
\[
U(g)\calB(\calO)U(g)^{-1}=\calB(g\calO),\quad g\in SO_0(1,d),
\]
where $U$ is the representation of the de Sitter group associated with
$\calA$.  $\calB$ is extended to any region as above.

Clearly $\calB$ satisfies the properties {\bf a)}, {\bf b)} and {\bf c)}, 
but for the cyclicity of $\Q$. By the Reeh-Schlieder theorem argument
\begin{equation}
\overline{(\vee_{g \in SO_0(d,1)}\calB (g\calO))\Q}= \overline{\calB 
(\calO)\Q} \quad \calO\in\tilde\calK\ ,
\end{equation}
where the bar denotes the closure, thus we have 
\begin{equation}\label{wedgecyclicity}
\overline{\calB(W)\Q}=\calH_{\calB}\equiv\overline{\calB(dS^d)\Q}
\end{equation}
for every $W\in\calW$, because the de Sitter group acts transitively
on $\calW$ and every double cone is contained in a wedge.  Thus
$\overline{\calB(D)\Omega}$ is independent of the region $D\subset
dS^2$ if $D$ contains a wedge.

Clearly $\calB$ acts on $\calH_{\calB}$ and we denote by $\calB_0$ its 
restriction to $\calH_{\calB}$. Note that $\calB_0$ is net 
satisfying all properties {\bf a)}, {\bf b)}, {\bf c)}, but not 
necessarily {\bf d)}. 

We shall say that a subnet $\calB$ is \emph{expected} (in $\calA$) if 
for every $\calO\in\calK$ there exists a vacuum preserving 
conditional expectation of $\calA(\calO)$ onto $\calB(\calO)$
\[
\varepsilon_{\calO}:\calA(\calO)\to\calB(\calO)
\]
such that
\[
\varepsilon_{\calO}\upharpoonright_{\calA(\calO_0)}=\varepsilon_{\calO_0}
\quad \calO_0\subset\calO \ , \calO_0,\calO\in\tilde\calK\ .
\]
It is easily seen that that $\varepsilon_{\calO}$ is given by
\[
\varepsilon_{\calO}(X)\Q = E_{\calO}X\Q, \quad X\in\calA(\calO)\ ,
\]
where $E_{\calO}$ is the orthogonal projection onto $\overline{\calB(\calO)\Q}$.
\begin{Lemma}\label{expadd}
    If $\calB$ is expected, then $\calB_0$ is weakly additive.
\end{Lemma}
\begin{proof}
    Let $\calO$ be a double cone and $W\supset\calO$ a wedge.  If
    $X\in\calA(\calO)$ we have
    \[
    \varepsilon_{\calO}(X)\Q = E_{\calO}X\Q =\varepsilon_{W}(X)\Q= 
    E_{W}X\Q = EX\Q \,
    \]
    where $E$ is the projection on $\calH_{\calB}$, hence
    \[
    \overline{\calB(\calO)\Q} = 
    \overline{\varepsilon_{\calO}(\calA(\calO))\Q}= 
    \overline{E\calA(\calO)\Q} = E\calH = \calH_{\calB}\ ,
    \]
    namely $\calB_0(\calO)$ is cyclic on $\Q$.  The statement then
    follows by Lemma \ref{wa}.
\end{proof}
The following Lemma is elementary, but emphasizes a property that 
holds on $dS^d$ but not on on $M^d$ and makes a qualitative difference 
in the subnet analysis in the two cases.

\begin{Lemma}\label{wedges}
Let $\calO$ be a double cone in $dS^d$. The the union of all wedges in 
$\calW_{\calO}\equiv\{W\in\calW : W\supset\calO\}$
has non-empty causal complement (it is the double cone antipodal to 
$\calO$).
\end{Lemma}
\begin{proof}
If $x\in dS^d$, let $\bar x$ denote its antipodal point. 
If $W$ is a wedge, then $W'$ is the antipodal of $W$, hence
$W$ contains $x$ iff $W'$ contains $\bar x$. If $\{W_i\}$ is a 
family of wedges, then 
\[
x\in\bigcap_i W_i \Leftrightarrow 
\bar x\in\bigcap_i W'_i = (\bigcup_i W_i)'\ .
\]
Thus if $\cap_i W_i =\calO$ the spacelike complement of $\cup_i W_i$ 
is the antipodal of $\calO$.
\end{proof}

\begin{Prop}\label{dual}
Let $\calA$ be a local net on $dS^d$ on a Hilbert space $\calH$, 
and $\calB$ a subnet. Setting 
$\hat\calB(\calO)=\cap_{W\in\calW_{\calO}}\calB(W)$, the following hold:
\begin{itemize}
\item[$(i)$]
$\hat\calB$ restricts to $\widehat{\calB_0}$ on $\calH_{\calB}$ (the dual 
net of $\calB_0$).
\item[$(ii)$]
$\calB_0$ is Haag dual iff $\calB=\hat\calB$.
\item[$(iii)$]
$\hat\calB$
is an expected subnet of $\hat\calA$. 
\item[$(iv)$]
$\calB$ is expected in $\hat\calA$ iff $\calB_0$ is Haag 
dual.
\end{itemize}
\end{Prop}
\begin{proof}
$(i)$: By Lemma \ref{dual} $D'\neq \emptyset$ where $D\equiv\cup\{W: 
W\in\calW_{\calO}\}$. Hence $\Omega$ is separating for 
$\calA(D)$, so the map $X\in\calB(D)\mapsto 
X\!\upharpoonright_{\calH_{\calB}}$ is an isomorphism between 
$\calB(D)$ on $\calH$ and $\calB(D)$ on $\calH_{\calB}$.
It follows that the operation $\cap_{W\in\calW_{\calO}}\calB(W)$ of taking 
intersection commutes with the restriction map.

$(ii)$: By Prop. \ref{Haag duality} $\calB$ is Haag dual iff 
$\widehat{\calB_0}=\calB_0$, thus iff $\calB=\hat\calB$ by the 
previous point.

$(iii)$: If $W$ is a wedge then by the geodesic KMS property and Takesaki 
theorem there exists a vacuum preserving conditional expectation 
$\varepsilon_W:\calA(W)\to\calB(W)$ such that
\[
\varepsilon_W(X)E=EXE, \quad X\in\calA(W),
\]
where $E$ is the orthogonal projection onto 
$\overline{\calB(W)\Q}=\calH_{\calB}$, (cf. \cite{CDR,L}). 

Let $X\in\hat\calA(\calO)$ and $W\in\calW_{\calO}$. Since $X\in\calA(W)$, 
we have $\varepsilon_W(X)\in\calB(W)$, so there exists $Y_W\in\calB(W)$ 
such that $Y_W E=EXE$. If $W_1$ is another wedge in $\calW_{\calO}$ 
then $Y_{W_1}\Omega =EX\Omega = Y_W\Omega$, thus 
$Y_{W_1} = Y_W$ because $\Omega$ is separating for 
$\calB(W_1)\vee\calB(W)$ by Lemma \ref{wedges} and Reeh-Schlieder 
theorem. Thus the operator $Y\equiv Y_W$ is independent of 
$W\in\calW_{\calO}$ and belongs to $\calB(W)$ for all wedges in $\calW_{\calO}$,
namely $Y\in\hat\calB(\calO)$. The map $X\mapsto Y$ is clearly a vacuum 
preserving conditional expectation from $\hat\calA(\calO)$ onto 
$\hat\calB(\calO)$.

$(iv)$: If $\calB_0$, then 
$\calB=\hat\calB$ by $(i)$ and $\calB$ is expected in $\hat\calA$ by $(iii)$. 

Conversely, assume that $\calB$ is expected in $\hat\calA$. We have 
\[
\hat\calB(\calO)=
\bigcap_{W\in\calW_{\calO}}\calB(W)\subset
\bigcap_{W\in\calW_{\calO}}\calA(W)=\hat\calA(\calO)\ ,
\]
so, if $X\in\hat\calB(\calO)$, then $X\in\hat\calA(\calO)$ and $EXE=XE$, 
namely $\varepsilon_{\calO}(X)=X$, so $X\in\calB(\calO)$. Thus 
$\hat\calB=\calB$.
\end{proof}

\section{Conformal Fields}

\subsection{Basics on the conformal structure}
\label{CFTondS}

It is a known fact that several interesting spacetimes can
be conformally embedded in the Einstein static universe, see
\cite{HE,BD}. We shall recall here some  embeddings and we begin 
with a discussion about conformal transformations.

\subsubsection{The conformal group and the conformal completion}

Two metrics on a manifold are said to belong to the same conformal
class if one is a multiple of the other by a strictly positive
function.  Given two semi-Riemannian manifolds $\man_{1}$, $\man_{2}$,
a \emph{local conformal map} is a triple $(\calD_{1},\calD_{2},T)$
where $\calD_{1}\subset\man_{1}$, $\calD_{2}\subset\man_{2}$ are open,
non-empty sets and $T:\calD_{1}\mapsto\calD_{2}$ is a diffeomorphism
which pulls back the metric on $\man_{2}$ to a metric in the same
conformal class as the original metric on $\man_{1}$.

With $\man$ a $d$-dimensional semi-Riemannian manifold, a 
\emph{conformal vector field} is a vector field $Z$ on $\man$ that 
satisfies the conformal Killing-Cartan
equation: there exists a function $f$ such that
\begin{equation}\label{CKe}
    \langle\mathbf{\nabla}_{Z}X,Y\rangle+
    \langle X,\mathbf{\nabla}_{Z}Y\rangle=
    f\langle X,Y\rangle,
\end{equation}
for all vector fields $X,Y,Z$.

Conformal vector fields form a Lie algebra (the exponentiate to local
one-pa\-ra\-me\-ter groups to local conformal maps, see below).  We shall
now assume $d\geq3$ (our discussion will motivate definitions also in
the $d=2$ case).  The dimension of the Lie algebra of the conformal
Killing vector fields is then finite and indeed lower or equal than
$(d+1)(d+2)/2$, the equality holding if and only if the manifold is
conformally flat, namely the metric tensor is equal to the flat one up
to a nonvanishing function \cite{TMP}.  Such Lie algebra is called the
\emph{conformal Lie algebra} of $\man$, and is denoted by
$\conf(\man)$.  Let us observe that such Lie algebra does not really
depend on the metric on $\man$, but only on the conformal class,
namely two metrics on $\man$ in the same conformal class give rise to
the same Lie algebra $\conf(\man)$.

Let us recall now that a Lie group $G$ \emph{acts locally} on a manifold $\man$
if there exists an open set $W\subset G\times \man$ and a $C^\infty$
map
\begin{gather}\label{(1.1)}
T: W \to \man\\
(g,x)\mapsto T_gx
\end{gather}
with the following properties:
\begin{itemize}
    \item[$(i)$] $\forall x\in \man$, $V_x\equiv \{g\in G:(g,x)\in
    W\}$ is an open connected neighborhood of the identity $e\in G$;
    \item[$(ii)$] $T_ex=x$, $\forall x\in \man$;
    \item[$(iii)$] If $(g,x)\in W$, then
    $V_{T_gx}=V_xg^{-1}$ and moreover for any $h\in G$ such
    that $hg\in V_x$ 
    $$T_h T_gx=T_{hg}x.$$
\end{itemize}

In general, a vector field satisfying equation $(\ref{CKe})$ gives
rise to a one-parameter family of (non-globally defined)
transformations that are local conformal mappings, namely to a local action 
of $\mathbb R$ on $\calM$ by means of local conformal maps, therefore
$\conf(\man)$ exponentiates to a (connected, simply connected) group
acting on $\man$ by local conformal mappings.  We shall call this Lie
group the \emph{local conformal group} of $\man$, and denote it by
$\CONF_{loc}(\man)$.

A manifold $\calM$ is \emph{conformally complete} if the elements of
$\CONF_{loc}(\man)$ are everywhere defined maps, i.e.
$\CONF_{loc}(\man)$ is contained in $\CONF(\man)$, the group of global
conformal transformations of $\calM$.  Obviously, in this case
$\CONF_{loc}(\man)$ is contained in the $\CONF_0(\man)$, the connected
component of the identity in $\CONF(\man)$.

\begin{Lemma}
    The stabilizer $H$ of a point $x$ under the action of the group
    $\CONF_{loc}(\man)$ is a closed subgroup.
\end{Lemma}

\begin{proof}
    Let's prove the group property.  Indeed if $g\in V_{x}$ stabilizes
    $x$ then $V_{x}$ is $g$-invariant, by $(iii)$ above.  Then, if
    $g,h\in V_{x}$ stabilize $x$, $h\in V_{x}=V_{x}g^{-1}$, namely
    $hg\in V_{x}$, and clearly $T_{hg}x=x$.  Now assume $g_{n}\to g$,
    $g_{n}\in V_{x}$ and $g_{n}x=x$.  Then there exists $n_{0}$ such
    that, for $n>n_{0}$, $g_{n}^{-1}g\in V_{x}$, therefore
    $g=g_{n}\cdot g_{n}^{-1}g\in V_{x}$, and $gx=x$ follows by
    continuity.
\end{proof}

Let us assume that $\CONF_{loc}(\man)$ acts transitively on $\man$. 
We may therefore identify $\man$ with an open subspace of the
homogeneous space $\widetilde{\man}=\CONF_{loc}(\man)/H$.  Clearly,
the Lie algebra $\conf(\widetilde{\man})$ coincides with the Lie
algebra $\conf(\man)$, and $\CONF_{loc}(\man)$ acts globally on
$\widetilde{\man}$.  Therefore the local conformal group of $\calM$
($=$ the local conformal group of $\widetilde\calM$) acts globally on
$\widetilde{\man}$, namely $\widetilde\man$ is conformally complete.

Let us note that in general the action of $\CONF_{loc}(\man)$ may be
non-effective on $\widetilde{\man}$, namely there may be non-identity
elements of $\CONF_{loc}(\man)$ acting trivially.  Therefore in
general $\CONF_{0}(\widetilde{\man})$ is a quotient of
$\CONF_{loc}(\man)$.

Now we come back to the case of a non conformally complete manifold
$\man$ on which $\CONF_{loc}(\man)$ acts transitively, and suppose
that there exists a discrete central subgroup $\Gamma$ of
$\CONF(\widetilde{\man})$ such that $\man$ is a fundamental domain for
$\Gamma$, namely $\cup_{\Gamma}\gamma\man$ is dense in 
$\widetilde{\man}$ and the
$\g\man$'s are disjoint.  Then $\widetilde{\man} / \Gamma$ is
conformally complete and $\man$ embeds densely in it.  In this case,
$\widetilde{\man} / \Gamma$ is denoted by $\overline{\man}$ and is
called the conformal completion of $\man$. (In the cases we shall 
consider, and possibly in all cases, the choice of $\Gamma$ is 
unique, thus the definition of $\overline{\man}$ does not depend on 
$\Gamma$).

We now summarize the construction of the conformal completion 
in the following diagram:

\[
\CD
\calM @>\text{conf. Killing v. fields}>>
\text{conf}(\calM)   @>\text{exponential}> 
>  \text{CONF}_{loc}(\calM)
\\ @ V \text{completion} V V 
& &@ V \text{(transitive case)} V H \, \text{stabilizer} V  \\  
\overline{\calM}
@=  
\tilde\calM/\Gamma
@<\Gamma \, \text{discrete central} < < 
\tilde\calM =\text{CONF}_{loc}(\calM)/H 
\endCD
\]

Clearly $\CONF(\overline{\man})$ acts on $\man$ by restriction.  Such action is
indeed quasi global \cite{BGL}, namely the open set
$$\{x\in \man:(g,x)\in W\}$$
is the complement of a meager set $S_g$, and the following
equation holds:
\begin{equation}\label{(1.2)}
    \lim_{x\to x_0}T_gx=\infty,\qquad g\in G,\quad x_0\in S_g\ ,
\end{equation}
where $x$ approaches $x_0$ out of $S_g$ and a point goes to infinity
when it is eventually out of any compact subset of $\man$.  It has
been proved in \cite{BGL} that any quasi-global action of a Lie group
$G$ on a manifold $\man$ gives rise to a unique $G$-completion, namely 
to a unique manifold $\overline{\man}$ in which $\man$ embeds densely 
and on which the action of $G$ is global.

Now $\CONF(\overline{\man})$ acts quasi-globally on $\man$, we shall
follow the standard usage in physics and call it the \emph{conformal
group} of $\man$, denoting it by $\Conf(\man)$.  Of course when $\man$
is conformally complete $\CONF(\man)=\Conf(\man)$.

In the $d=2$ case, the Lie algebra of conformal Killing vector fields 
is infinite-dimensional. The above discussion goes through by 
considering a finite-dimensional sub-Lie group. In the Minkowski 
spacetime (an in conformally related spacetimes, see Sect. \ref{NetsdSM})
this is the Lie algebra of the group generated by the 
Poincar\'e group and the ray inversion map.

Analogous considerations can be made for isometries, namely by 
replacing conformal Killing vector fields by Killing vector 
fields which are
obtained setting $f=0$ in equation $(\ref{CKe})$. These gives rise to the
local one-parameter groups with values in $\Iso_{loc}(\man)$, the 
local isometry group.  If a Lorentzian manifold is geodesically
complete then $\Iso_{loc}(\man)$ acts globally on it (cf.  e.g.
\cite{KN}).

\subsubsection{The embedding of $M^{d}$}

Einstein static universe $E^d=\mathbb R \times S^d$ may be defined as
the cylinder with radius 1 around the time axis in the
$d+1$-dimensional Minkowski spacetime $M^{d+1}$, equipped with the
induced metric.  

Denoting the coordinates in $M^{d+1}$ by $(t_{E},\mathbf{x}_{E},w_{E})$ 
and the coordinates in $M^{d}$ by $(t_{M},\mathbf{x}_{M})$, we consider 
the embedding
\begin{equation}
    \left\{
    \aligned
    \mathbf{x}_{E}&=(\eta+r_{M}^{2})^{-1/2}\mathbf{x}_{M}\\
    w_{E}&=\text{sgn}(\eta)(\eta+r_{M}^{2})^{-1/2}\\
    t_{E}&=\arctan(t_{M}+r_{M})+\arctan(t_{M}-r_{M})
    \endaligned
    \right.
\end{equation}
which maps the $d$-dimensional Minkowski space into the Einstein
universe, where we have set $r_{M}=|\mathbf{x}_{M}|$,
$\eta=\frac{1}{2}(1+t_{M}^{2}-r_{M}^{2})$.

If we now use the cylindrical coordinates
$(t_{E},\boldsymbol{\theta}_{E},\psi_{E})$ in $M^{d+1}$ to describe $E^{d}$,
and the cylindrical coordinates $(t_{M},r_{M},\boldsymbol{\theta}_{M})$ in
$M^{d}$, we get
\begin{itemize}
    \item the metric tensor of $E^{d}$ is $\text{d}s_{E}^2 =
    \text{d}t_{E}^2 - \text{d}\psi_{E}^2 -
    \sin^2\psi_{E}\text{d}\Q(\boldsymbol{\theta}_{E})^2$, where
    $\text{d}\Q(\boldsymbol{\theta}_{E})^2$ denotes the metric tensor of the
    $(d-2)$-dimensional unit sphere; 
    \item the metric tensor for $M^{4}$ is $\text{d}s_{M}^2 =
    \text{d}t_{M}^2 - \text{d}r_{M}^2 - r_{M}^2\text{d}
    \Q(\boldsymbol{\theta}_{M})^2$;
    \item the embedding above can be written as
    \begin{equation}
	\left\{
	\aligned
	\boldsymbol{\theta}_{E}&=\boldsymbol{\theta}_{M}\\
	\psi_{E}&=\arctan(t_{M}+r_{M})-\arctan(t_{M}-r_{M})\\
	t_{E}&=\arctan(t_{M}+r_{M})+\arctan(t_{M}-r_{M}).
	\endaligned
	\right.
    \end{equation}
\end{itemize}

A simple calculation shows that the metric tensor on $M^{d}$ is pulled
back to the following metric on $E^{d}$:
$$
\text{d}s^2 =\frac14\sec^{2}\left(\frac{t_{E}+\psi_{E}}{2}\right)
\sec^{2}\left(\frac{t_{E}-\psi_{E}}{2}\right) \text{d}s_{E}^2,
$$
showing that the embedding is conformal and that the image of $M^d$ is
the ``double cone'' of $E^d$ given by
\begin{equation}\label{dc}
-\pi < t_{E}\pm\psi_{E} < \pi.
\end{equation}

\begin{rem}
    In the two dimensional case $\psi$ is the only angle coordinate, 
    hence it ranges from $-\pi$ to $\pi$, and the previous inequality 
    is indeed drawn as a double cone. In higher dimension, 
    $\psi\in[0,\pi]$, however the inequality still describes a double 
    cone in $E^{d}$ with center $(t_{0},\mathbf{v}_{0})$:
    \begin{equation}
	\{(t,\mathbf{v})\in\R\times S^{d-1}:
	|t-t_{0}|+\mathrm{d}(\mathbf{v},\mathbf{v}_{0})<\pi\},
    \end{equation}
    where $t_{0}=0$, $\mathbf{v}_{0}$ is the point 
    $\mathbf{x}_{E}=0$, $w_{E}=1$, and $\mathrm{d}(\cdot,\cdot)$ 
    denotes the geodesic distance in $E^{d}$.
\end{rem}

The conformal Lie algebra of $M^{d}$ is $o(d,2)$.  If $d\geq3$, the
quotient of the universal covering of $SO_{0}(d,2)$ by the stabilizer
of a point is $E^{d}$.  However the action of the universal covering
of $SO_{0}(d,2)$ is not effective, since there is a $\mathbb{Z}_{2}$
component in the center acting trivially on $E^{d}$.  The
corresponding quotient is (the identity component of) the conformal
group of $E^{d}$.

Indeed let us now consider the map $\gamma$ in $E^{d}=\mathbb R\times
S^{d-1}$ given by $\gamma : (t_{E},\mathbf{v}) \mapsto
(t_{E}+\pi,-\mathbf{v})$, where $\mathbf{v}\mapsto -\mathbf{v}$ is the
antipodal map.  It is easy to see that $\gamma$ belongs to
$\Conf(E^{d})$ and the ``double cone'' above is a fundamental domain
for the corresponding action of $\mathbb Z$ on $E^{d}$.  Therefore the
quotient is the conformal completion $\overline{M^{d}}$ of $M^{d}$,
which is usually called the Dirac-Weyl compactification of $M^{d}$. 
Since $\Gamma^{2}$ is central $\Conf_{0}(E^{d})$, the quotient
$SO_{0}(d,2) = \Conf(E^{d})/2\mathbb{Z}$ acts on $\overline{M^{d}}$. 
If $d$ is even, such action is not effective on $\overline{M^{d}}$,
and the (quasi-global) conformal group of $M^{d}$ is $PSO_{0}(d,2)$. 
If $d$ is odd, the action is effective, and $\Conf(M^{d}) =
SO_{0}(d,2)$.

If $d=2$, the conformal group is infinite dimensional, however we
shall still set $\conf(M^{2})=o(2,2)$.  Moreover, $E^{2}$ is not
simply connected, indeed the procedure outlined above would give as
$\widetilde{M}^{2}$ the universal covering of $E^{2}$.  However,
$E^{2}$ is the only globally hyperbolic covering of $\overline{M}^{2}$
where the image of $M^{2}$ has empty space-like complement.  As we
shall see below, this condition is necessary in order to lift a
conformal net on $E^{2}$ to a local net, therefore we shall write
$\widetilde{M}^{d}=E^{d}$ when $d=2$ too.

\subsubsection{The embedding of $dS^{d}$}

The de Sitter space $dS^{d}$ (of radius $\rho$) may be described in
terms of the coordinates $(\tau, \boldsymbol{\theta}_{S},\psi_{S})$, where
$\tau$ varies in $(0,\pi)$, $(\boldsymbol{\theta}_{S}$ are spherical
coordinates in $S^{d-2}$ and $\psi_{S}$ varies in $[0,\pi]$, such that
$(\boldsymbol{\theta}_{S},\psi_{S})$ are spherical coordinates in $S^{d-1}$. 
Then the embedding of $dS^{d}$ in $M^{d+1}$ is 
\begin{equation}
    \left\{
    \aligned
    t&=-\rho\cot \tau\\
    \mathbf{x}&=\rho\cdot(\sin\tau)^{-1}\mathbf{v}(\boldsymbol{\theta}_{S},\psi_{S}),
    \endaligned
    \right.
\end{equation}
where $\mathbf{v}(\boldsymbol{\theta}_{S},\psi_{S})$ denotes a point in
$S^{d-1}$ expressed in terms of spherical coordinates.  In terms of
these coordinates the metric tensor is
$$
\text{d}s^2 =\frac{\rho^{2}}{\sin^{2}\tau} ( \text{d}\tau^2
-\text{d}\psi_{S}^2 - \sin^2(\psi_{S})\text{d}\Q(\boldsymbol{\theta}_{S})^2).
$$

Therefore the embedding of $dS^{d}$ in $E^{d}$
\begin{equation}
    \left\{
    \aligned
    \f_{E}&=\f_{S}\\
    \boldsymbol{\theta}_{E}&=\boldsymbol{\theta}_{S}\\
    \psi_{E}&=\psi_{S}\\
    t_{E}&=\tau
    \endaligned
    \right.
\end{equation}
is conformal and maps $dS^{d}$ to the ``rectangle'' of $E^{d}$
\begin{equation}\label{rect}
    \{(t_{E},\mathbf{x}_{E},w_{E}):
    |\mathbf{x}_{E}|^{2}+w_{E}^{2}=1,0<t_{E}<\pi\}.
\end{equation}

Again, the rectangle is a is a fundamental domain for the action of
$\mathbb Z$ on $E^{d}$ induced by $\Gamma$.  Therefore
$\Conf(M^{d})=\Conf(dS^{d})$ and $\overline{M^{d}}=\overline{dS^{d}}$,
$\Conf(dS^{d})$ acting quasiglobally on $dS^{d}$.  When $d=2$ we
define $\conf(dS^{2})=o(2,2)$.  Let us note that, opposite to the
$M^{2}$ case, the homogeneous space given by the quotient of
$\CONF_{loc}(dS^{2})$ by the stabilizer of a point is exactly $E^{2}$,
not its covering.

\subsubsection{The conformal steady-state universe}

The intersection of the conformal images of $M^{d}$ and $dS^{d}$ in 
$E^{d}$ is the steady-state universe. Composing the previous maps we 
may therefore obtain a conformal map from the subspace $\{t_{M}>0\}$ 
in the Minkowski space to the steady state subspace of the de Sitter space.

The map can be written as a map from $\{t_{M}>0\}$ in $M^{d}$ to
$M^{d+1}$, with range $\{(t,\mathbf{x},w)\in
M^{d+1}:-t^{2}+|\mathbf{x}|^{2}+w^{2}=\rho^{2}, w>t\}$:
\begin{equation}
    \left\{
    \aligned
    t&=-\rho\cdot\frac{t_{M}^{2}-|\mathbf{x}_{M}|^{2}-1}{2t_{M}}\\
    \mathbf{x}&=\rho\cdot\frac{\mathbf{x}_{M}}{t_{M}}\\
    w&=-\rho\cdot\frac{t_{M}^{2}-|\mathbf{x}_{M}|^{2}+1}{2t_{M}}.
    \endaligned
    \right.
\end{equation}

The image of the steady state universe in $E^{d}$ is not a fundamental 
domain for some $\Gamma$, therefore there is no conformally complete 
manifold in which it embeds densely.

Let us note for further reference that time translations for $t_{M}>0$
are mapped to endomorphisms of the steady-state universe, and that the
(incomplete) time-like geodesic $\{\mathbf{x}_{M}=0,t_{M}>0\}$ is
mapped to the de Sitter-complete geodesic
\begin{equation}
    \left\{
    \aligned
    t&=-\rho\cdot\frac{t_{M}^{2}-1}{2t_{M}}\\
    \mathbf{x}&=0\\
    w&=-\rho\cdot\frac{t_{M}^{2}+1}{2t_{M}}.
    \endaligned
    \right.
\end{equation}
\medskip
\begin{tabular}{l l}
    \epsfbox{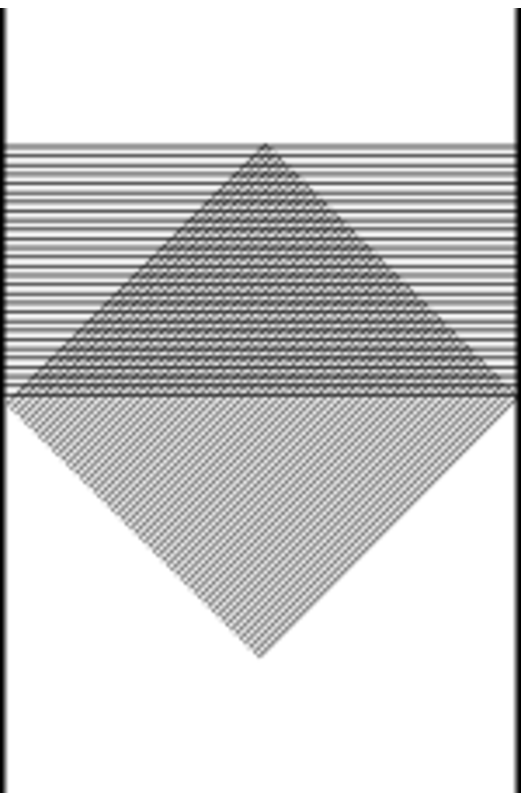} &
    \vbox{
    \hbox{\hsize=2.7in \vbox{\lineskip=4pt\noindent Fig.~3.  
    The embeddings of Minkowski space, de~Sitter space,
    and steady state universe in Einstein universe.}}
    \hbox{\vbox{\vskip1.2in}}
    \hbox{\hsize=2.5in\epsfbox{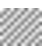}\hskip0.2in
    \vbox{\lineskip=4pt\noindent Minkowski space.}}
    \hbox{\vbox{\vskip0.2in}}
    \hbox{\hsize=2.5in\epsfbox{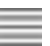}\hskip0.2in
    \vbox{\lineskip=4pt\noindent de Sitter space.}}
    \hbox{\vbox{\vskip 0.2in}}
    \hbox{\hsize=2.5in\epsfbox{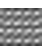}\hskip0.2in
    \vbox{\lineskip=4pt\noindent Steady state universe.}}
    }
\end{tabular}

\subsection{Conformal nets on de Sitter and Minkowski spacetimes}
\label{NetsdSM}

Let $\man$ be a spacetime on which the local
action of the conformal group is transitive.

A net $\calA$ of local algebras on $\man$ is \emph{conformal} if
there exists a unitary local representations $U$ of
$\CONF_{loc}(\man)$ acting covariantly: for each fixed double cone
$\calO$ there exists a neighbourhood $\calU$ of the identity in
$\CONF_{loc}(\man)$ such that $g\calO\subset \man$ for all $g\in\calU$
and
\[
U(g)\calA(\calO)U(g)^{-1} = \calA(g\calO),\quad \forall g\in\calU \ .
\]

\begin{Prop}\label{preconformal lift}
    A conformal net $\calA$ on $\man$ lifts to a net
    $\widetilde{\calA}$ on $\widetilde{\man}$ and the local
    representation $U$ lifts to a true representation $\widetilde{U}$
    of $\CONF_{loc}(\man)$ under which $\widetilde{\calA}$ is
    covariant.
\end{Prop}

\begin{proof}
    The results follow as in \cite{BGL}: for any region $\calO_1\equiv
    g\calO\subset\widetilde{\man}$, $\calO\subset\man$, $g\in
    \CONF_{loc}(\man)$, we set $\widetilde{\calA}(g\calO) =
    \widetilde{U}(g) \calA(\calO) \widetilde{U}(g^{-1})$, and observe
    that $\calA(\calO_1)$ is well defined since $\CONF_{loc}(\man)$ is
    simply connected.  Then we extend the net $\widetilde{\calA}$ on
    all $\widetilde{\man}$ by additivity.
\end{proof}

Let us notice that we did not assume $\calA$ to be local, namely that
commutativity at spacelike distance is satisfied.  In particular we
did not prove that $\widetilde{\calA}$ is local.  This was proved in
\cite{BGL} for local nets on the Minkowski space, and the proof easily
extends to a spacetime $\man$ where the conformal group acts quasi
globally and whose conformal completion is the Dirac-Weyl space, as is
the case for the de Sitter space.  We shall prove a more general
result here.

We say that a (local) unitary representation of $\Conf(E^{d})$ has 
positive energy if the generator of the one-parameter group of time 
translations on $E^{d}$ is positive.

Let us denote by $\calK$ the set of double cones of $E^d$ (the
definition is analogous as in the de Sitter case), and by
$\Lambda_{\calO}$ the one-parameter group of conformal transformation
of $E^d$, that can be defined by requiring that $\Lambda_{W}$ is the
boost one parameter group associated with $W$ if $W$ is a wedge of
$M^{d}$ embedded in $E^{d}$, and
$\Lambda_{\calO}(t)=g\Lambda_{W}(t)g^{-1}$ if $\calO\in\calK$ and $g$
is a conformal transformation such that $gW=\calO$.

We shall say that a net on $E^{d}$ satisfies the \emph{double cone KMS property} 
if, for any $\calO\in\calK$, $(\Q,\ \cdot\ \Q)$ is a KMS state on the 
algebra associated with $\calO$ w.r.t. the evolution implemented by
$U(\Lambda_{\calO}(\cdot))$.

\begin{Thm}\label{conformal lift}
    Let $\man$ be a spacetime s.t. $\widetilde{\man}=E^{d}$.  Then a
    local conformal net $\calA$ on $\man$ with positive energy lifts
    to a local net $\tilde\calA$ on $E^{d}$ which is covariant under
    the (orientation-preserving) conformal group $\CONF_{+}(E^{d})$.
    
    $\tilde\calA$ satisfies Haag duality and the double cone KMS property.
\end{Thm}

\begin{proof}
    By the above proposition we get a net on $\widetilde{\man}$ which
    is covariant under the universal covering of $SO(d,2)$.  Then
    modular unitaries associated with double cones act geometrically,
    as in \cite{BGL} Lemma 2.1.  Now we fix two causally disjoint
    double cones $\calO,\calO_{1}\subset \man$.  Then if
    $\Lambda_{\calO}$ is the one parameter group of conformal
    transformations corresponding to the (rescaled) modular group of
    $\calA(\calO)$, we have that $\calA(\calO)$ commutes with
    $\widetilde{\calA}(\Lambda_{\calO}(t)\calO_{1})$ for any $t$.  Let 
    us assume for the moment that $d>2$. Then
     $\Lambda_{\calO}$ leaves globally invariant the
    spacelike complement $\calO'$ of $\calO$ in $\widetilde{\man}$,
    indeed its action is implemented by the modular group of
    $\widetilde{\calA}(\calO')$ at $-t$.  Therefore the algebra
    \[
    \bigvee_{t\in\mathbb R}\widetilde{\calA}(\Lambda_{\calO}(t)\calO_{1})
    \]
    is a subalgebra of $\widetilde{\calA}(\calO')$, is globally stable
    under the action of $\Delta_{\calO'}^{it}$, commutes with
    $\calA(\calO)$ and is cyclic for the vacuum.  By the Takesaki
    theorem the subalgebra indeed coincides with
    $\widetilde{\calA}(\calO')$.  This implies that the net is local
    by covariance.  One then proves the geometric action of $J$, thus
    extending the representation to conformal transformations which does not
    preserve the time orientation.  $\widetilde{U}$ is a
    representation of the conformal group of $E^{d}$ rather than of
    its simply connected two-covering by a spin and statistics
    argument (cf.  \cite{gl:95,Ku1,GLRV1}).  The last properties are
    proved as in \cite{BGL}.  In the low-dimensional case we may
    assume that $\calO_{1}$ is ``on the right'' of $\calO$ and
    $\cup_{t\in\mathbb R}\Lambda_{\calO}(t)\calO_{1}=\calO^{R}$, where
    $\calO^{R}$ is the closest smallest region on the right of $\calO$
    which is globally invariant under $\Lambda_{\calO}(t)$.  As
    before, we prove that $\calA(\calO^{R})=\calA(\calO)'$.  Now there
    exists a suitable conformal rotation whose lift $R(t)$ to $E^{d}$
    satisfies $R(\pi)\calO=\calO^{R}$,
    $R(\pi)\calO^{R}=(\calO^{R})^{R}$, and so on.  Therefore,
    $\calA(R(2\pi)\calO)=\calA(\calO)''=\calA(\calO)$, namely the net
    actually lives on $E^{d}$.  The rest of the proof goes on as
    before.
\end{proof}

\begin{rem}
    In the proof above, we proved in particular that, when $d\leq2$,
    the extension of the net satisfying locality necessarily lives on
    $E^{d}$, and not on its universal covering.
\end{rem}

Besides the Minkowski space and the de Sitter space, Th. 
\ref{conformal lift} applies to the Robertson-Walker space $RW^{d}$,
to the Rindler wedge and many others.  In particular, there is a
bijection between isomorphism classes:
\begin{equation*}
\boxed{\text{local conformal nets on}\ M^d
\rightleftharpoons
\text{local conformal nets on}\ dS^d}
\end{equation*}
In the following theorem we describe what the positive energy
condition on $M^d$ becomes on $dS^d$ under this correspondence.
\begin{Thm}\label{equivalence of nets}
    There is a natural correspondence between isomorphism classes of
    \begin{itemize}
	\item[$(i)$] Local conformal nets on $M^d$ with positive energy;
	\item [$(ii)$] Local conformal nets on $dS^d$ with the KMS property 
	for geodesic observers;
	\item [$(iii)$] Local conformal nets on $E^d$ with positive energy.
    \end{itemize}
    Here positive energy on $E^d=\mathbb R\times S^{d-1}$ means that the 
    one-parameter group of time translations (on $\mathbb R$) is 
    implemented by a unitary group with positive generator.
\end{Thm}
\begin{proof} 
    $(i)\Leftrightarrow(iii)$: Let us note that the universal covering
    $\widetilde{SO}(d,2)$ of $SO_{0}(d,2)$ is also the universal
    covering of $SL(2,\R)$.  Since the covariance unitary
    representation $U$ of $\widetilde{SO}(d,2)$ is the same, it
    suffices to show that the two one-parameter unitary subgroup of
    $U$ in question both have or not have positive generators.

    Let's consider the group generated by time translation, dilations
    and ray inversion in $M^d$.  This group is isomorphic to
    $PSL(2,\mathbb R)$ and acts on time axis of $M^d$.  $U$ restricts
    to a unitary representation of $PSL(2,\mathbb R)$ thus, by a well
    known fact (see e.g. \cite{K}, positivity of time translations on
    $M^d$ is equivalent to positivity of conformal rotations (the
    generator corresponding to the rotation subgroup of $PSL(2,\mathbb
    R)$ is positive).  Now the above rotation group provides the time
    translations on $E^d$, hence the positivity of the corresponding
    one-parameter subgroup of $\widetilde{SO}(d,2)$ is a consequence
    of the mentioned equivalence of positive energy conditions for
    unitary representations of (the universal covering of)
    $PSL(2,\mathbb R)$.

    $(iii)\Rightarrow(ii)$ has been proved above.

    $(ii)\Leftrightarrow(i)$: It is known that a Poincar\'e covariant
    net for which the vacuum is KMS for the algebra of a wedge acted
    upon by the boosts satisfies the spectrum condition, see e.g.
    \cite{BS,BGL2}.  Since geodesic KMS property for a net on $dS^{d}$
    is equivalent to the KMS property for any wedge for the
    corresponding net on $M^{d}$, we get the thesis.
\end{proof}

If $\calO$ is a double cone with vertices $x$ and $y$, call 
$B$ a basis of $\calO$ if $B$ is the part of a Cauchy surface 
contained in $\calO$ and the closure of $B$ contains the points 
lightlike to both $x$ and $y$. 
We shall say that a net $\calA$ satisfies the \emph{local time-slice 
property} if for any double cone $\calO$ if 
\[
\calA(\underline{\calO})=\calA(\calO)
\]
with $\calO\in\calK$ and $\underline{\calO}\subset\calO$ an open slice around a 
basis $B$ of $\calO$, namely a tubular neighbourhood of $B$ contained 
in $\calO$ (thus $\calO =\underline{\calO}''$).
Note that, by an iteration/compactness argument, for the local timelike slice 
property to hold it is enough to assume 
$\calA(\underline{\calO})=\calA(\calO)$ where $\underline{\calO}$ is 
obtained by $\calO$ by removing arbitrarily small neighbourhoods of
the vertices of $\calO$ (by using additivity).
\begin{Cor}\label{time-slice}
Let $\calA$ be  a conformal net on $dS^d$. $\calA$ is Haag dual iff it 
satisfies the local time-slice property.
\end{Cor}
\begin{proof}
With $dS^d$ is embedded in $E^d$ as above and $\calO$ a double cone 
in $dS^d$, we have $\calA(\calO)' = \calA(\calO'_E)$ where $\calO'_E$ 
is the causal complement of $\calO$ in $E^d$. Thus $\calA$ is Haag 
dual on $dS^d$ iff $\calA(\calO') = \calA(\calO'_E)$, where 
$\calO'=\calO'_E\cap dS^d$ is the causal complement of $\calO$ in $dS^d$.
Now $\calO'_E$ is a double cone in $E^d$ and $\calO'$ is a timelike 
slice for $\calO'$, so Haag 
duality on $dS^d$ is satisfied iff the time-slice holds for 
$\calO'_E$. We can now map, by a conformal transformation, $\calO'_E$ 
to any other double cone contained in $dS^d$, thus the time-slice 
property holds on $dS^d$ iff it holds for $\calO'_E$.
\end{proof}

Thus, under a general assumption (local time-slice property), all 
conformal nets on $dS^2$ are Haag dual. One should compare this with 
the Minkowski spacetime case, where Haag duality for conformal nets 
is equivalent to a strong additivity requirement:
removing a point from the basis $B$ of $\calO$ we have 
$\calA(\calO)=\calA(B\setminus\{\text{pt}\})''$ \cite{HL}. As a 
consequence, if two conformal nets on $M^d$ and $dS^d$ are conformally related 
as above, then
\[
\boxed{\text{Haag duality on $M^d$}\implies \text{Haag duality on $dS^d$}}
\]
but the converse is not true.

\subsection{Modular covariance and the maximal conformal subnet}

We have shown that on spacetimes that can be conformally embedded in
$E^{d}$, a local, locally conformal net can be lifted to a local,
globally conformal net on $E^{d}$ with the double cone KMS property,
namely to a net for which the modular groups of double cones have a
geometric action.  Indeed a converse is true.  Assume we have a
spacetime $\man$ such that $\widetilde\man=E^{d}$.  We shall say that
a subregion $\calO$ of $\man$ is a double cone if it can be
conformally identified with a double cone in $E^{d}$.  Given a net
$\calA$ on the double cones of $\man$ acting on a Hilbert space with a
given vector $\Omega$, such that, for any double cone
$\calO\subseteq\man$, $\Omega$ is cyclic and separating for
$\calA(\calO)$, we shall consider the following property for the
algebra $\calA(\calO)$:

\begin{itemize}
    \item {\emph{Local modular covariance:}} for every double cone 
    $\tilde\calO\subset\calO$, we have
    $$
    \Delta_{\calO}^{it} \calA(\tilde\calO) \Delta_{\calO}^{it} =
    \calA(\Lambda_{\calO}(-2\pi t)\tilde\calO).
    $$
\end{itemize}

Local modular covariance was introduced in \cite{Gu} under the name of
weak modular covariance, where it was proven that weak modular
covariance for wedges plus essential duality is equivalent to modular
covariance, hence reconstructs the Poincar\'e covariant
representation, for nets on the Minkowski space.

\begin{Thm}\label{localreconstruction}
    Let $\calO_{0}$ be a spacetime which can be conformally identified
    with a double cone in $E^{d}$, and assume we are given a net
    $\calO \to \calA(\calO)$ of local algebras, $\calO \subset
    \calO_{0}$, acting on a Hilbert space with a given vector
    $\Omega$, such that, for any double cone $\calO \subseteq
    \calO_{0}$, $\Omega$ is cyclic and separating for $\calA(\calO)$
    and the local modular covariance property holds.  Then the net
    extends to a conformal net on (the universal covering of) 
    $E^{d}$. If $\calA$ is local, then the extended net is indeed a 
    local conformal net on $E^{d}$.
\end{Thm}

The proof requires some steps.  

We first construct ``half-sided modular translations''.  Let us
identify $\calO_{0}$ with a future cone in $M^{d}$, and denote by $v
\mapsto \tau^{+} (v)$ the subgroup of the conformal group isomorphic
to $\R^{d}$ consisting of $M^{d}$ translations, in such a way that
when $v$ is a causal future-pointing vector $\tau^{+}(v)$ implements
endomorphisms of $\calO_{0}$.  These transformations can be seen as
conformal translations which fix the upper vertex of $\calO_{0}$.  In
the same way we get a family $v\mapsto\tau^{-}(v)$ of conformal
translations fixing the lower vertex of $\calO_{0}$, and such that
$\tau^{-}(v)$ implements endomorphisms of $\calO_{0}$ when $v$ is a
causal past-pointing vector.
    
For any causal future-pointing vector $v$, we may implement the
translation $t\mapsto\tau^{+}(tv)$ by a one-parameter unitary group
$T^{+}(tv)$ with positive generator {\it a la Wiesbrock}.  Borchers
relations are satisfied: $\Delta_{ \calO_{0} }^{it} T^{+}(v) \Delta_{
\calO_{0} }^{ -it } = T^{+}(e^{-2\pi t}v)$.
    
Translations $T^{-}(v)$, for causal (past-pointing) vector $v$, are
constructed analogously.

\begin{Lemma}
    The $T^{\pm}$ translations associated with $\calO_{0}$ act
    geometrically on subregions, whenever it makes sense:
    \begin{equation}\label{Tgeoaction}
	\Ad T^{\pm}(v)\calA(\calO)=\calA(\tau^{\pm}(v)\calO),\quad 
	{\mathrm{if}\ } \tau^{\pm}(v)\calO\subset\calO_{0}.
    \end{equation}
\end{Lemma}

\begin{proof}
    First assume $\calO$ is compactly contained in $\calO_{0}$ and the
    translation ``goes inside'', namely it is of the form
    $\tau^{+}(v)$ with $v$ a causal future-pointing vector or
    $\tau^{-}(v)$ with $v$ a causal past-pointing vector.  Then there
    exists an $\eps>0$ such that $\tau^{+}(\eps
    v)\calO\subset\calO_{0}$.  Therefore, by Borchers relations,
    $T^{+}(\eps(e^{-2\pi t}-1)v)=\Delta_{\calO_{0}}^{it}
    \Delta_{\tau^{+}(\eps v)\calO_{0}}^{-it}$, and the thesis follows.
    
    Then, by additivity, one can remove the hypothesis that $\calO$ is
    compactly contained in $\calO_{0}$.  Indeed, by local modular
    covariance, for any $\calO\subset\calO_{0}$, the von Neumann
    algebra generated by the local algebras associated with compactly
    contained subregions of $\calO$ is globally stable for $\Delta_{
    \calO }^{it}$, therefore, by Takesaki Theorem, it coincides with
    $\calA(\calO)$.
    
    We have proved that (\ref{Tgeoaction}) holds for $\tau^{+}(v)$
    whenever $v$ is a causal future-pointing vector, hence, applying
    $\Ad T^{+}(-v)$, one gets $\Ad T^{+}(-v) \calA(\calO) = \calA
    (\tau^{+}(-v) \calO)$ whenever $\tau^{+}(-v) \calO \subset
    \calO_{0}$.  The thesis follows.
\end{proof}

\begin{Lemma}
    $T^{+}$ is indeed a representation of $\R^{d}$, and the same holds
    for $T^{-}$.  They act geometrically on subregions, whenever it
    makes sense.
\end{Lemma}

\begin{proof}
    First we prove, as in \cite{Gu}, that $[T^{+}(v),T^{+}(w)]=0$.  By
    the previous point, the multiplicative commutator
    \begin{equation}\label{1.16}
	c(s,t):=T^{+}(-sv)T^{+}(-tw)T^{+}(sv)T^{+}(tw)
    \end{equation}
    has a geometric action, hence stabilizes, the algebras
    $\calA(\calO)$, for $s,t\geq0$. Therefore it commutes with
    $\De_{\calO}^{it}$ and with the translations themselves. With 
    simple manipulations we get $c(s,t) = c(-s,-t) = c(-s,t)^{*} = 
    c(s,-t)^{*}$, namely $c(s,t)$ commutes with translations for any 
    $s,t$, hence $T^{+}(sv),T^{+}(tw)$ generate a central extension 
    of $\R^{2}$. By positivity of the generators the commutator has to 
    vanish.
    
    In an analogous way one shows that $c(t):= T^{+} (-t(v+w)) T^{+} (tv)
    T^{+} (tw)$ is central, hence is a one-parameter group, and by 
    Borchers relations $c(\lambda t)=c(t)$ for any positive 
    $\lambda$, namely $c(t)=1$. The relations for $T^{-}$ and the 
    geometric action follows as before.
\end{proof}

Now we construct the group $\calG$.  For any $\calO \subseteq
\calO_{0}$, define $\calG (\calO)$ as the group generated by
$$\{\Delta_{\tilde\calO}^{it}:\tilde\calO \subseteq \calO\}.$$

\begin{Lemma}\label{calG}
    $\calG(\calO)$ is independent of $\calO$.
\end{Lemma}

\begin{proof}
    Let us note that $\calG(\tau^{\pm}(v)\calO)$ is a subgroup of 
    $\calG(\calO)$ and clearly 
    contains $T^{\pm}(v)$, hence coincides with $\calG(\calO)$. 
    Repeating this argument we get that $\calG(\calO)$ does not depend 
    on $\calO$. 
\end{proof}

We shall denote this group simply by $\calG$.  Let us note that
$\calG$ is generated by a finite number of one-parameter groups:
setting $\calO_{k} = \tau^{+} (v_{k}) \calO_{0}$, $k=1,\dots,d$,
$\calO_{k+d}=\tau^{-}(v_{k})\calO_{0}$, $k=1,\dots,d$, the
one-parameter groups $\Delta_{\calO_{k}}^{it}$, $k=0,\dots 2d$
generates all translations $T^{\pm}(v)$, hence $\calG$ by covariance.
    
Then we construct the central extension.  The one-parameter groups
$\Lambda_{k}$ $k=1,\dots,d$, generate the conformal group
$SO_{0}(d,2)$.  Pick $(d+1)(d+2)/2$ functions $g_{i}(t)$ with values
in $SO_{0}(d,2)$, each given by a product of $\Lambda_{k}$'s , such
that the Lie algebra elements $g_{i}'(0)$ form a basis for $so(d,2)$. 
Since $(d+1)(d+2)/2\geq 2d+1$ one may assume that $g_{i}(t) =
\Lambda_{ \calO_{k-1} }(-2\pi t)$, $i=1,\dots 2d+1$.  Then the map
$F(\mathbf{t}) = g_{1}(t_{1}) \cdots g_{n}(t_{n})$ $n=(d+1)(d+2)/2$,
is a local diffeomorphism from $\R^{n}$ to the conformal group.  Now
use the identification $\Lambda_{\calO}(-2\pi t)\leftrightarrow
\Delta_{\calO}^{it}$ to get a map $G$ from $\R^{n}$ to the group
$\calG=\calG(\calO_{0}\dots\calO_{2d})$ generated by the
$\Delta_{\calO_{k}}^{it}$, $k=0,\dots 2d$, and finally obtain a map
$H=G\cdot F^{-1}$ from a neighborhood $\calV$ of the identity in
$SO_{0}(d,2)$, to $\calG$.  Observe that $\Ad H(g) \calA(\calO) =
\calA(g\calO)$ whenever each step makes sense.

\begin{Lemma}
    The inverse of the map $H$ gives rise to a homomorphism from $\calG$
    to $SO_{0}(d,2)$ which is indeed a central extension.
\end{Lemma}

\begin{proof}
    First we show that $H$ is a local homomorphism to $\calG/\calZ$, 
    $\calZ$ denoting the center of $\calG$.
    
    Choose a region $\tilde\calO$ compactly contained in $\calO_{0}$. 
    Now, possibly restricting $\calV$, one may assume that
    $g\tilde\calO\subset\calO$ for any $g\in\calV$.  As a consequence,
    if $g,h,gh\in\calV$, then $H(gh)^{*}H(g)H(h)$ implements an
    automorphism of $\calA(\calO)$, for $\calO\subset\tilde\calO$,
    namely commutes with the corresponding modular groups, hence is in
    the center of $\calG$.
    
    Now we extend the map $H$ to a homomorphism from the universal
    covering $\widetilde{SO}(d,2)$ of $SO_{0}(d,2)$ to $\calG/\calZ$,
    and observe that since all normal subgroups of $\widetilde{SO}
    (d,2)$ are central, we get an isomorphism from a suitable covering
    $\tilde\calC$ of $SO_{0}(d,2)$ to $\calG/\calZ$.  The inverse
    gives rise to a homomorphism from $\calG$ to $SO_{0}(d,2)$ which is
    indeed a central extension.
\end{proof}

\begin{proof}[Proof of Theorem \ref{localreconstruction}]
    The arguments in \cite{BGL2} show that the extension is weak Lie
    type, hence gives rise to a strongly continuous representation $U$
    of $\widetilde{SO}(d,2)$.  Such representation acts geometrically
    on the algebras $\calA(\calO)$ whenever it makes sense, therefore,
    by Proposition \ref{preconformal lift} we get a CFT on (the universal
    covering of) $E^{d}$.  If $\calA$ is local, the extension is
    indeed a local net on $dS^{d}$ by Theorem \ref{conformal lift}.
\end{proof}

In the following Corollary we characterize conformal theories in terms
of local modular covariance.
\begin{Cor}\label{confchar}
    Let $\man$ be a spacetime for which $\widetilde\man=E^{d}$. Then
    there is a natural correspondence between 
\begin{itemize}
\item Local conformal nets on $E^d$ with positive energy;
\item Local  nets on $\man$ with local modular covariance for 
double cones.
\end{itemize}
\end{Cor}

\begin{proof} 
    Assume we are given a local net $\calA$ on $\man$ satisfying local
    modular covariance for double cones.  For any double cone $\calO
    \subset \man$, Theorem \ref{localreconstruction} gives a local
    conformal net $\widetilde \calA_{ \calO}$ on $E^{d}$, based on the
    restriction of $\calA$ to $\calO$.
    
    Now embed $\man$ in $E^{d}$, and observe that, by Lemma
    \ref{calG}, if $\calO_{1} \subset \calO_{2} \subset dS^{d}$ the
    two nets $\widetilde \calA_{ \calO_{i}}$, $i=1,2$, on $E^{d}$
    coincide.  From this we easily get that all nets $\widetilde
    \calA_{ \calO}$ based on $\calA|_{\calO}$ coincide, hence their
    restriction to $\man$ coincides with $\calA$.  The converse
    implication follows by Theorem \ref{conformal lift}.
\end{proof}

Note that the unitary representation of the conformal group is unique
\cite{BGL} because it is generated by the unitary modular groups
associated with double cones (cf. Thm.  \ref{localreconstruction}).

A further consequence of Corollary \ref{confchar} is \emph{additivity}
for a conformal net $\calA$: if $\calO,\calO_i$ are double cones and
$\calO\subset\cup_i\calO_i$, then $\calA(\calO)\subset\vee_i
\calA(\calO_i)$.  This can be proved by the argument in \cite{FJ}. 

\bigskip 

We now return to the de Sitter spacetime $dS^d$, with any dimension
$d$.  Let $\calA$ be a local net on $dS^d$ and $\calB$ be a subnet of
$\calA$.  We shall say that $\calB$ is a \emph{conformal subnet} if
its restriction $\calB_0$ to $\calH_{\calB}$ is a conformal net.  Now,
given any local net $\calA$ and $\calO\in\calK$, we set
\begin{equation}\label{confsubnet}
\calC(\calO)=\{X\in\calA(\calO):
\Delta_{\calO_0}^{it}X\Delta_{\calO_0}^{-it} \in
\calA(\Lambda_{\calO_0}(-2\pi\rho t)\calO), \forall \calO_0\in\calK,
\calO_{0}\supset\calO\}.
\end{equation}
It is immediate to check that $\calC(\calO)$ is a von Neumann 
subalgebra of $\calA(\calO)$. Moreover $\calC$ is covariant w.r.t. the 
unitary representation of $SO_0(d,1)$ because if $X\in\calC(\calO)$, then 
$\Ad U(g)X\in\calA(g\calO)$ and, for any $\calO_{0}\supset 
g\calO$,
\begin{multline}
\Ad\Delta_{\calO_0}^{it}U(g)X=
\Ad U(g)\Delta^{it}_{g^{-1}\calO_0}X\\
\in\Ad U(g)\calA(\Lambda_{g^{-1}\calO_0}(-2\pi\rho t)\calO)
=\calA(\Lambda_{\calO_0}(-2\pi\rho t)g\calO),
\end{multline}
namely $U(g)XU(g)^{-1}\in\calC(g\calO)$.

Finally $\calC$ is isotonic, thus $\calC$ is a subnet of $\calA$.
\begin{Thm}\label{maximal conformal}
    A local net $\calA$ on $dS^d$ has a unique maximal conformal
    expected subnet $\calC$.  It is given by eq.  (\ref{confsubnet}).
\end{Thm}
\begin{proof}
Let $\calB$ be a conformal expected subnet of $\calA$.  Then
$\calB_{0}$ is weakly additive by Lemma \ref{expadd}, hence satisfies
the Reeh-Schlieder property by Lemma \ref{wa}.  So the projection $E$
onto $\calH_{\calB}$ implements all the expectations
$\varepsilon_{\calO}$ and commutes with all $\Delta_{\calO}$ by
Takesaki theorem.  By Corollary \ref{confchar}, local modular
covariance is satisfied for $\calB(\calO)$, hence if $X$ is an element
of the algebra $\calB(\calO)$ it belongs to $\calC(\calO)$.

Thus we have only to show that the subnet $\calC$ is conformal and
expected.  Clearly $\Delta_{ \calO }^{it} \calC (\calO) \Delta_{ \calO
}^{-it} = \calC (\calO)$, thus $\calC$ is expected by Takesaki
theorem.  Also, by construction, local modular covariance holds true,
so $\calC$ is conformal by Corollary  \ref{confchar}.
\end{proof}

\section{The Dethermalization Effect} 
In the flat Minkowski spacetime, the world line of an inertial
particle is a causal line.  The corresponding evolution on a quantum
field in the vacuum state is implemented by a one-parameter
translation unitary group whose infinitesimal generator, ``energy'', is
positive.  A uniformly accelerated observer feels a
thermalization (Unruh effect): its orbit is the orbit of a
one-parameter group of pure Lorentz transformation that, on the
quantum field, is implemented by a one-parameter automorphism group of
the von Neumann algebra of the corresponding wedge that satisfies the
KMS thermal equilibrium condition at Hawking temperature.

On the other hand, if we consider an inertial observer on the de
Sitter spacetime, its world line is the orbit of a boost and it is
already thermalized, in the vacuum quantum field state, at Gibbons-Hawking
temperature.  Our aim is to seek for a different evolution on the de
Sitter spacetime with respect to which the vacuum is dethermalized,
namely becomes a ground state, an effect opposite to the Unruh
thermalization.

\subsection{General evolutions}
Let us recall a recent proposal of (quasi-)covariant dynamics for 
not necessarily inertial observers \cite{BMS2}.
Our presentation, though strictly paralleling the one in \cite{BMS2}, 
will differ in some respects. Our description is in fact strictly 
local, therefore local conformal transformations will play the central role.
The dynamics will consist of propagators describing the time 
evolution as seen by the observer, the main requirement being 
that the rest frame for the observer is irrotational.

Let us consider a (not necessarily parametrized by
proper time) observer in a given
spacetime $\man$, namely a timelike, future pointing $C^{1}$ curve
$\g:t\in (-a,a)\mapsto \g_{t}\in\man$.

Then we look for a local evolution for the observer $\g$, namely a
family  of maps $\lambda_{t}$ from $\man$ to $\man$ such that satisfy 
the following physical requirements 
\begin{enumerate}
	\item[$\bullet$] $\lambda_{t}\gamma_{0}=\g_{t}$, $t\in(-a,a)$.  
	\item[$\bullet$] Given $x_0 \in\calM$, 
	for each $y_0$ in some neighborhood of $x_{0}$, the 
	events $\lambda_{t}(y_0)$, $t\in (-a,a)$, describe, potentially, 
	the worldline of some material particle.  This worldline is either
	disjoint from the observer's worldline or coincides with it.
	\item[$\bullet$]  For a suitable $y_0$ spacelike to $x_0$, the axis of 
	a gyroscope carried by the observer at the space-time point 
	$\lambda_{t}(x_0)$ points towards the point $\lambda_{t}(y_0)$ at 
	all times $t$.
\end{enumerate}

As observed in \cite{BMS2}, the previous conditions only depend on the 
conformal structure of the manifold. Therefore we will specify 
$\lambda_{t}$ to be a local conformal transformation of $\man$, or, 
more precisely, $\lambda$ to be a curve in $\Conf(\overline{\man})$. 

In this way the notion of local evolution only depends on the
conformal class, namely if two metrics belong to the same conformal
class they give rise to the same notion of local evolution.  From the
mathematical point of view, the above requests mean that the range of $\gamma$ is
an orbit of $\lambda$, and that for any $t\in(-a,a)$,
$(\lambda_{t})_{*}$ (the differential of the transformation
$\lambda_t:\man\mapsto\man$) maps orthogonal frames in
$T_{\g_{0}}\man$ to orthogonal frames in $T_{\g_{t}}\man$, in such a
way that a tangent vector to the curve $\g$ at $t=0$ is mapped to a
tangent vector to the curve $\gamma$ at the point $t$, and that every
orthogonal vector $v$ to $\g$ at $t=0$ evolves without rotating to
vectors orthogonal to $\gamma$, as we will explain.

If we now fix a metric $g$ in the conformal class, we can choose the
proper time parametrization, and then look for a curve $\lambda\in
\Iso(\man,g)$, the isometry group of $\man$, namely for a local isometric
evolution on $\gamma$.

In this way orthonormal frames evolve to orthonormal frames. 
Recalling the notion of Fermi-Walker transport (cf.  \cite{SW}), we
may reformulate the conditions for a local isometric evolution on
$\gamma$ as follows:

\begin{enumerate}
    \item[$E_1$:] $\lambda$ is a curve in $\Iso(\man,g)$.
    \item[$E_2$:] $\lambda_{t}\gamma_{0}=\g_{t}$, $t\in(-a,a)$.  
    \item [$E_3$:] $(\lambda_{t})_{*}$ is the Fermi-Walker transport
    along the curve $\gamma$.
\end{enumerate}

Clearly a local isometric evolution on $\gamma$ does not exist in
general however, if it exists, it is unique.

\begin{Prop}\label{uniqueIsoFW}
    Assume $\lambda_{t},\lambda'_{t}$, $t\in(-a,a)$, satisfy
    properties $E_1$, $E_2$, $E_3$ for a given observer $\g$.  Then
    $\lambda_{t}$ coincides with $\lambda_{t}'$ on a suitable
    neighborhood of $\g_{0}$.
\end{Prop}

\begin{proof}
	By assumption, $\lambda_{t}^{-1}\cdot \lambda'_{t}$ is a local
	isometry fixing the point $x_{0}$ whose differential is the
	identity on $T_{x_{0}}\man$.  Then $\lambda_{t}^{-1}\cdot
	\lambda'_{t}$ acts identically on any geodesic at $x_{0}$,
	hence coincides with the identity on the injectivity radius 
	neighborhood.
\end{proof}

Concerning the existence problem, let us first consider a geodesic
observer.  In this case the Fermi-Walker transport coincides with the
parallel transport.  Let us recall that a (semi-) Riemannian manifold
is {\it symmetric} if for any $p\in \man$ there exists an involutive
isometry $\sigma_{p}$ such that $p$ is an isolated fixed point.  It is
easy to see that de Sitter, Minkowski, and Einstein spacetimes are
symmetric.

\begin{Prop}\label{geodesicFW}
    \item{$(i)$} If $\gamma$ is a geodesic observer, a local isometric
    evolution is indeed a one-parameter group of isometries.
    \item{$(ii)$} If $\man$ is symmetric, a local isometric evolution 
    exists for any geodesic.
\end{Prop}

\begin{proof}
    $(i)$ First we show that a local isometric evolution for a geodesic
    $\gamma$ satisfies $\lambda_{t}\cdot \lambda_{t} = \lambda_{2t}$. 
    Indeed, since $\lambda_{t}$ is an isometry,
    $\lambda_{t}\gamma_{s}$, $0\leq s\leq t$ describes a geodesic, and
    since $(\lambda_{t})_{*}\gamma'_{0}=\gamma'_{t}$, it describes the
    geodesic $\gamma_{t+s}$, $0\leq s\leq t$.  As a consequence,
    $\lambda_{2t}$ implements the parallel transport on $\gamma$ from
    $T_{\gamma_{0}}\man$ to $T_{\gamma_{t}}\man$. By the uniqueness 
    proved in Proposition \ref{uniqueIsoFW} we get the statement.
    
    Now we observe that the previous property implies
    $\lambda_{t}\cdot \lambda_{s} = \lambda_{t+s}$ whenever $s/t$ 
    is rational, hence, by continuity, for any $s$ and $t$.
    
    $(ii)$ Since $\gamma$ is geodesic, the Fermi-Walker transport
    coincides with the parallel transport (cf.  \cite{SW}).  On a
    symmetric manifold, the existence of isometries implementing the
    parallel transport is a known fact, see e.g. \cite{Boo}, Thm
    8.7.
\end{proof}

We now study the case of a generic observer.  Assume $\lambda$ is
a $C^{1}$ one-parameter family of local diffeomorphisms of $\man$
and denote by $L_{t}$ the vector field given by $L_{t}(\lambda_{t}(x))
= \frac{d}{ds}\lambda_{s}(x)|_{s=t}$.  Assume then that the
$x$-derivatives of $L_{t}(x)$ are jointly continuous, namely that the
map
\begin{equation}\label{diffcond}
    (t,x,v)\in\R\times T\man\mapsto (\nabla_{v}L_{t})(x)
\end{equation}
is continuous.

\begin{Lemma} 
     Let $\gamma$ be an orbit of $\lambda$: $\lambda_{t}\gamma_{0} =
     \g_{t}$, and let $X$ be a $\lambda$-invariant vector field on
     $\gamma$: $(\lambda_{t})_{*}X(\gamma_{0})=X(\gamma_{t})$.  Then,
     at the point $\gamma_{t}$, the covariant derivative of $X$ on the
     curve $\gamma$ satisfies
    \begin{equation}\label{difgeoformula}
	\nabla_{\gamma^{*}}X =
	\nabla_{X}L_{t}.
    \end{equation}
\end{Lemma}

\begin{proof}
    Since $X$ is invariant under $\lambda$, the commutator $[X,L_{t}]$
    vanishes at the point $\gamma_{t}$. This fact can be proved via 
    a simple computation, where two derivatives should be exchanged. 
    Condition (\ref{diffcond}) ensures that Schwartz Lemma applies.
    
    Then the symmetry of the Levi-Civita connection implies
    $$
    \nabla_{L_{t}}X = \nabla_{X}L_{t}
    $$
    at the point $\gamma_{t}$.  Since by definition $L_{t}(\gamma_{t})
    = \frac{d}{ds}\lambda_{s}(\gamma_{0})|_{s=t} =
    \frac{d}{ds}\gamma_{s}|_{s=t}$, we get the thesis.
\end{proof}

The existence of a local isometric evolution for any observer has been
proved in \cite{BMS2} for the de Sitter metric.  Property $(iii)$ of
the following theorem gives as an extension of this fact.

\begin{Thm}
    Let $\gamma$ be an observer in $\man$. The following hold:
    \begin{itemize}
	\item[$(i)$] There exists a local isometric evolution
	$\lambda$ on $\gamma$ satisfying condition
	(\ref{diffcond}) {\it iff} for every $t$, $\gamma'_{t}$ extends
	locally to a Killing vector field $L_{t}$ satisfying
	(\ref{diffcond}) and $(\nabla_{v}L_{t}(\gamma_{t}),w)=0$ for
	every vectors $v,w$ in the rest space of $\gamma_{t}$. 
	\item[$(ii)$] The existence of a local isometric evolution for
	any geodesic observer is equivalent to the existence of a
	local isometric evolution for any observer.  
	\item[$(iii)$] If $\man$ is symmetric, a local isometric
	evolution exists for every observer.
    \end{itemize}
\end{Thm}

\begin{proof}
    $(i)$ A curve $\lambda$ in $\Iso(\man)$ satisfying
    $E_{2}$ on $\gamma$ gives rise, by derivation, to a one-parameter
    family of Killing vector fields $L_{t}$ defined by:
    $L_{t}(\lambda_{t}(x)) = \frac{d}{ds}\lambda_{s}(x)|_{s=t}$.
    Clearly $L_{t}$ satisfies $L_{t}(\gamma_{t})=\gamma'_{t}$.

    Conversely a curve $L_{t}$ of Killing vector fields verifying
    $L_{t}(\gamma_{t})=\gamma'_{t}$ gives
    rise to a curve of local isometries via the equations

	\begin{align*}
	    \lambda_{0}(x)&=x\\
	    \left.\frac{d\lambda_{s}(x)}{ds}\right|_{s=t}
	    &=L_{t}(\lambda_{t}(x)).
	\end{align*}

    Clearly $\frac{d}{ds}\lambda_{s}(\gamma_{t})|_{s=t} =
    L_{t}(\gamma_t) = \gamma'_{t}$, hence $\lambda_{t}(\gamma_{0}) =
    \gamma_{t}$, namely condition $E_{2}$. 
    
    By condition $E_{2}$, $(\lambda_{t})_{*}$ maps vectors tangent to
    $\gamma$ to vectors tangent to $\gamma$, hence, being isometric,
    preserves the rest frame for $\gamma$.  Therefore it implements
    the Fermi-Walker transport if and only if tangent rest vectors
    evolve irrotationally, namely {\it iff} the Fermi derivative
    $\mathbf{F} _{\gamma^{*}}X=0$ on $\gamma_{t}$ for any
    $\lambda$-invariant vector field $X$ in the rest space of
    $\gamma$.  According to \cite{SW}, Prop.  2.2.1, if $P$ denotes
    the projection on the rest space, this is equivalent to
    $P\nabla_{\gamma^{*}}X=0$.  By equation (\ref{difgeoformula}),
    this means that
    \begin{equation}\label{localcond}	
	P\nabla_{v}L_{t}(\gamma_{t})=0,\quad v\in PT_{\gamma_{t}}\man,\quad\forall t,
    \end{equation}
    which is our thesis.

    $(ii)$ Assume the existence of a local isometric evolution for any
    geodesic observer.  By Proposition \ref{geodesicFW}, $L_{t}$ does
    not depend on $t$, hence condition (\ref{diffcond}) is trivially
    satisfied.  Then, reasoning as in $(i)$ and taking into account
    that the Fermi-derivative for a geodesic observer is indeed the
    Levi-Civita connection, we get $\nabla_{w}L(x)=0$ for any $x$ in
    the geodesic, $w\in T_{x}\man$.  Namely, the existence of a local
    isometric evolution for any geodesic observer is equivalent to the
    following: for any $(x,v)\in T\man$, there exists a vector field
    $H=H_{x,v}$ defined in a neighborhood $\calU$ of $x$, such that, if
    $\gamma$ is the geodesic determined by $(x,v)$, $H$ satisfies
    \begin{align*}
	&(V, \nabla_{H} W)(x)=(\nabla_{H} V, W)(x),\quad x\in \calU,\\
	&\nabla_{w}L(x)=0,\quad w\in T_{x}\man,
	\quad x=\gamma(s),\quad |s|<\eps,\\
	&L(\gamma_{s})=\gamma'_{s},\quad  |s|<\eps
    \end{align*}
    where $\gamma(s)\subset\calU$ for any $|s|<\eps$.  Since any such
    $H_{x,v}$ would determine a local isometric evolution for
    $\gamma$, Proposition \ref{uniqueIsoFW} imply uniqueness.  Hence
    the existence of a local isometric evolution for any geodesic
    observer is equivalent to the existence and uniqueness of a local
    solution for the system above.  Let us remember that the solutions
    of the first equation (the Killing equation) form a finite
    dimensional space $\calV$, therefore existence and uniqueness can
    be reformulated as the existence and uniqueness for the
    finite-dimensional linear system given by the last two equations,
    with $L\in\calV$.  Clearly, both the linear operator and the
    coefficients depend smoothly on $(x,v)$ if the manifold (and the
    Riemannian metric) is smooth.  Therefore, for any (continuous)
    curve $\gamma$, the one-parameter family of Killing fields $L_{t}
    = H_{\gamma_{t},\gamma'_{t}}$ satisfies conditions
    (\ref{diffcond}) and (\ref{localcond}), namely, by point $(i)$,
    the existence of a local isometric evolution for any observer.
        
    $(iii)$ Immediately follows by Proposition \ref{geodesicFW} and
    point $(ii)$.
\end{proof}

\subsection{Dethermalization for conformal fields}

Besides the geometric question of existence of the curve $t\mapsto
\lambda_{t}\in\Conf(\man)$, there is a second existence problem if we
want to describe the local dynamics in quantum field theory.  Indeed,
it is not obvious that the local maps $\lambda_{t}$ are unitarily
implemented, or give rise to automorphisms of the net.  This is
clearly the case of a conformally covariant theory, but not the
general case.

The previous discussion on local evolutions shows that the evolution 
may change if we replace the original metric with another metric in 
the same conformal class.  We shall show that, with a suitable choice 
of the new metric, the original observer will become an inertial 
observer in a (locally) flat spacetime.  Therefore, in a conformal 
quantum field theory, the local evolution will be implemented by a 
one-parameter group with positive generator w.r.t.  which the vacuum 
state is a ground state.  

We mention the analysis contained in \cite{CD}, 
where the authors classify the global conformal 
vacua on a conformally flat spacetime in terms of the global timelike 
Killing vector fields. In the Minkowski spacetime there is only one 
global timelike Killing vector field, while for other ``small'' spacetimes one may 
have two nonequivalent Killing vector fields, as is the case of the 
Rindler wedge subregion where the boost flow is also timelike. The 
two Killing vector fields give rise to  different vacua, and the vacuum for the  
Minkowskian Killing vector field is thermalized w.r.t. the second 
Killing evolution.  Our construction represents a converse to this
procedure: starting with $dS^d$, where the global Killing vector field 
is unique and the de Sitter vacuum is thermalized, we restrict to a smaller 
spacetime where a global dethermalizing conformal Killing flow exists.

From a classical point of view then, the dethermalization is realized 
by replacing the original dynamics with a new `conformal' dynamics. 

Let us note that such a change of the dynamics implies in particular a
change in the time parametrization.  Of course the absence of a
preferred proper time parametrization occurs if the conformal structure
alone is considered.

As we shall see in the next sections, the evolutions $\lambda$
will give rise only to a quasi-covariant dynamics in the sense of
\cite{BMS2} for general (non conformal) quantum fields.

As seen in Subsection \ref{CFTondS}, there exists a conformal
diffeomorphism $\Psi$ between the steady-state universe subspace $\calN$
of $dS^{d}$ containing a given complete causal geodesics $\g$ and the
semispace $M^{d}_{+}=\{(\mathbf{x},t)\in M^{d}, t>0\}$ in the
Minkowski space, mapping $\g$ to a causal geodesics $\tilde{\g}$. 
However $\tilde{\g}$ is not complete, and can be identified with
the half-line $\{\mathbf x=0, t>0\}$ in the timelike case, and with
the half-line $\{x_{1}=t, x_{i}=0, i>1,t>0\}$ in the lightlike case. 
Therefore we get the following.

\begin{Prop}
    If we replace the metric on $\calN$ with the pull back via $\Psi$ of
    the flat metric on $M^{d}_{+}$, there exists a local evolution
    $\mu_{t}$, $t>0$, from $\calN$ into itself, given by the pull
    back of the time translations.
\end{Prop}
\begin{Thm}
    Let $\calA$ be a conformal net on $dS^{d}$ and $W$ a wedge
    causally generated by a geodesic observer $\g$.  Then:
\begin{itemize}
    \item[$(a)$] The local isometric evolution $\lambda$ corresponding
    to the de Sitter metric is indeed global, there exists one-parameter
    unitary group $U$ on the Hilbert space implementing $\lambda$ and
    the vacuum is a thermal state at the Gibbons-Hawking temperature
    w.r.t. $U$.
    
    \item[$(b)$] The local isometric evolution $\mu$
    corresponding to the flat metric is unitarily implemented, namely
    there exists a one-parameter unitary group $V$ on the Hilbert
    space such $V(t)$ implements $\mu_t$ for $t>0$, and the
    vacuum is a ground state w.r.t. $V$.  If we extend the net
    $\calA$ to a conformal net $\tilde\calA$ on the static Einstein
    universe, then $V(t)$ acts covariantly on $\tilde\calA$ for every
    $t\in\mathbb R$.
    \end{itemize}
\end{Thm}

\begin{proof}
    The first statement is simply a reformulation of assumption {\bf
    c)} in Section \ref{sec2}.  Concerning the second statement, note
    that $\mu$ extends to a global flow on $E^d$ which is
    implemented by a one parameter group $V$ with positive
    generator.  The thesis is then immediate.
\end{proof}

\medskip

\begin{tabular}{l l}
    \epsfbox{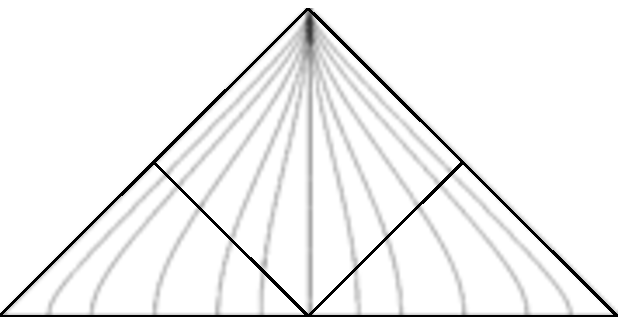} & \vbox{ \hbox{\hsize=2.1in
    \vbox{\lineskip=4pt\noindent Fig.~2.  The flow lines of the 
    isometric evolution $\lambda$ in the wedge contained in the 
    steady-state universe.}}}
\end{tabular}

\bigskip

\begin{tabular}{l l}
    \epsfbox{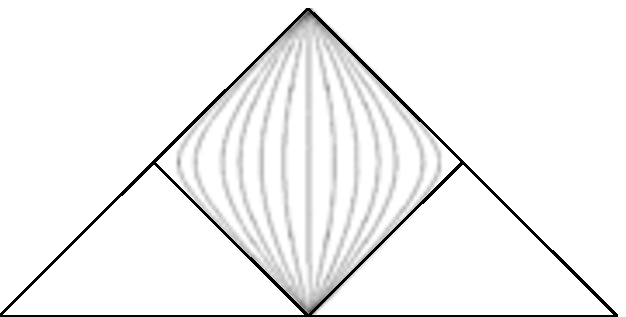} & \vbox{ \hbox{\hsize=2.1in
    \vbox{\lineskip=4pt\noindent Fig.~3.  The flow lines of the 
    dethermalizing evolution $\mu$ in the steady-state universe.}}}
\end{tabular}

\medskip

\begin{rem}
    In the conformal case, the dethermalizing evolution is not unique. 
    In fact we may identify $dS^{d}$ with a rectangle in the Einstein
    universe (cf.  eq.  (\ref{rect})), and then consider the corresponding
    metric on it.  Again, the new evolution, which is given by time
    translations in $E^{d}$, is dethermalized.
\end{rem}

\subsection{Dethermalization with noncommutative flows} 

As anticipated, we construct here a quasi-covariant dynamics 
corresponding to the geometric dynamics described above, showing that 
the vacuum vector becomes a ground state w.r.t. this dynamics.

Our flow will be noncommutative in the sense it give a noncommutative 
dynamical system, indeed it is a flow on a quantum algebra of 
observables, although it will retain a partial geometric action.

We begin with a no-go result. 

\begin{Prop}
Let $U$ be a non-trivial unitary representation of $SO_0(d,1)$, 
$d\geq 2$, and $u$ the associated infinitesimal representation of the Lie 
algebra $so(d,1)$. The following are 
equivalent:
\begin{itemize}
\item[$(i)$] There exists a non-zero $L\in so(d,1)$ such that $u(L)$ is a 
positive or negative operator;
\item[$(ii)$] $d=2$ and $U$ is the direct sum 
of irreducible representations that are either the identity or belong
to the discrete series of $SO_0(2,1)$ ($\simeq PSL(2,\mathbb R)$).
In this case $L$ belongs to the cone generated by the 
translation generators.
\end{itemize}
\end{Prop}
\begin{proof}
The set $\mathfrak P$ of $L\in so(d,1)$ such that $u(L)$ is a 
positive operator is a convex cone of $so(d,1)$, which is globally 
stable under the adjoint action of $SO_0(d,1)$, and 
$\mathfrak P\cap -\mathfrak P=0$ because $so(d,1)$ is a simple Lie 
algebra.

Now every element $L\in so(d,1)$ can be written as a sum $L= R+K$, 
where $R\in so(d)$ and $K$ is a the generator of a boost one-parameter subgroup.

Let then $L$ belong to $\mathfrak P$ and assume $d>2$.
We can then choose a rotation $r\in SO(d)$ such that Ad$r(K)=-K$.
Set $R'\equiv \Ad r(R)\in so(d)$.  Since 
$L'\equiv \Ad r(L)\in\mathfrak P$, the element
\[
L+L'= R+K + \Ad r(R)+\Ad r(K)= R + R'
\]
belongs to $\mathfrak P$ and to $so(d)$, so it is enough to show that
$\mathfrak P\cap so(d)=\{0\}$. Indeed if $R''\in so(d)$, $d>2$, we 
can choose a rotation $r$ such that $\Ad r(R'')=-R''$, thus 
$R''=0$ if $R''\in\mathfrak P$. 

We thus conclude that $d=2$.  Now every non-zero $L\in SO_0(2,1)
\simeq PSL(2,\mathbb R)$ is (conjugate to) the generator of either a
boost, or translation, or rotation one-parameter group.  If $L$ is a
boost generator, then $L$ is conjugate to $-L$ as above, thus
$L\notin\mathfrak P$.  The positivity of $u(L)$, $L$ a translation
generator, is equivalent to the positivity of $u(L)$, $L$ a rotation
generator (see e.g. \cite{K}) and is equivalent to $U$ to be a
direct sum of representations in the discrete series of $U$ and,
possibly, to the identity \cite{Lang}.
\end{proof}

The next corollary states that the existence of a dethermalized
covariant one-parameter dynamics is possible only if $d=2$ and implies
conformal covariance.

\begin{Cor}
    Given a local net $\calA$ on the de Sitter space, assume there is
    a one parameter group in $SO_0(d,1)$ which has positive generator
    in the covariance representation.  Then $\calA$ is conformally
    covariant.
\end{Cor}

\begin{proof}
    Assume the net is not conformally covariant.  By the Proposition
    above, this implies $d=2$.  Then, Corollary \ref{2d+E} shows that
    positive energy representations in the two-dimensional case imply
    conformal covariance.
\end{proof}

Now we turn to a geodesic observer $\gamma$, and denote by $W$ the
wedge generated by the complete geodesic, by $\calN$ the steady-state
universe containing $W$, by $\lambda$ the Killing flow
corresponding to the geodesic $\gamma$, by $\mu_{t}$, $t>0$,
the conformal evolution of $\calN$ described above.  Let us observe
that the time is reparametrized, namely $\tilde{\gamma}_{t}=\gamma_{\log
t}=\mu_{t-1}\g_{0}$.  We also denote by $R$ the spacetime
reflection mapping $W$ to its spacelike complement $W'$.

\begin{Thm}
    Let $\calA$ be a net of local algebras on the de 
    Sitter spacetime. Then there exists a unique one-parameter unitary 
    group $V$ with the following properties: 
    \begin{itemize} 
	\item[$(i)$]  $\Q$ is a ground state w.r.t. $V$;
	\item[$(ii)$] $V$ implements a quasi covariant dynamics for the
	regions $\mu_t(W)$, $t\geq0$, namely
	$V(t)\calA(W)V(-t)=\calA(\mu_tW)$, $t\geq0$;
	\item[$(iii)$] Partial localization for negative times:
	$V(-t)\calA(W)V(t)=\calA(R\mu_tW)'$, $t\geq0$.
    \end{itemize}
\end{Thm}

\begin{proof}
    By the geodesic KMS property, we get that 
    $\calA(\mu_1(W))\subset\calA(W)$ is a half-sided modular 
    inclusion. Therefore the theorem of Wiesbrock \cite{W} gives a 
    one-parameter group $V$ with positive generator such that 
    $V(1)\calA(W)V(-1)=\calA(\mu_1(W))$ and satisfying the 
    Borchers commutation relations
    \begin{align*}
	U(\lambda_{t})V(s)U(\lambda_{-t})=V(e^{t}s)\\
	U(R)V(s)U(R)=V(-s). 
    \end{align*}
    Then $(i)$ is obvious, and the above relations  give
    \begin{multline*}
    V(t)\calA(W)V(-t)= U(\lambda_{\log t})V(1)U(\lambda_{-\log t})
    \calA(W) U(\lambda_{\log t})V(-1)U(\lambda_{-\log t})\\
    =\calA(\mu_tW), t\geq0,
    \end{multline*}
    namely $(ii)$. Property $(iii)$ follows in an analogous way.
    
    The uniqueness now follows by the uniqueness for one-parameter groups 
with Borchers property \cite{Bo1} in the following lemma.
\end{proof}

\begin{rem}\label{localization}
    By the above theorem we have the following localization properties
    for the noncommutative flow $\Ad V$: 

    {$(i)$} If $\calL$ is a region contained in the steady state
    universe subregion $\calN$ of $dS$ and $\calL=\m_{s}W$ for some
    $s\geq0$, then $\m_{t}\calL\subset dS^{d}$ if and only if
    $t\in[-s,+\infty)$ and, for such $t$, $\Ad  V(t)
    \calA(\calL)=\calA(\m_{t}\calL)$.  Analogously, considering
    $\m_{t}$ as a global transformation on $E^{d}$ acting on $dS^{d}$
    by restriction, using the geometric action of $J$, and Borchers
    commutation relations $JV(t)J=V(-t)$, if $\calL$ is a region in
    the complement of $\calN$ and $\calL=\m_{-s}W'$ for some $s\geq0$,
    then $\m_{t}\calL\subset dS$ if and only if $t\in(-\infty,s]$ and,
    for such $t$, $\Ad  V(t) \calA(\calL)=\calA(\m_{t}\calL)$.

    {$(ii)$} If $\calO$ is a double cone contained in $\calN$,
    there exists $s>0$ such that, for any $s'\geq s$,
    $\calO\subset(\m_{-s'}W')'$.  Therefore, for any $t\in\R$,
    \begin{equation}
        \Ad V(t)\calA(\calO)\subset
        \begin{cases}
            \calA(\m_{t-s}W')'& \text{if}\  t-s\leq0\\
            \calA(\m_{t-s}W)& \text{if}\  t-s\geq0.
        \end{cases}
    \end{equation}
    Assuming Haag duality on $dS^d$ we then get
    \begin{equation}
        \Ad V(t)\calA(\calO)\subset
        \calA(\m_{t-s}W\cap dS^{d})
    \end{equation}
    for any double cone $\calO\subset\calN$; note that $\m_{t-s}W\cap 
    dS^{d}$ has non empty spacelike complement in $dS^d$.
    Analogous localization properties hold if $\calO$ is contained in 
    $dS^d\setminus\calN$.
    
    Localization results for any double cone $\calO\subset dS^d$ would 
    then follow by a form of strong additivity.
    
    Let us remark that more stringent localization properties would
    indeed imply a complete geometrical action \cite{Ku2}.
\end{rem}

\begin{Lemma}
    Let $\calP$ be a von Neumann algebra on a Hilbert space $\calH$ 
    with cyclic and separating vector $\Omega\in\calH$. Let $V_1$ and 
    $V_2$ be $\Omega$-fixing one-parameter unitary groups on $\calH$ 
    such that $V_k(t)\calP V_k(-t)\subset\calP$, $t\geq 0$, ($k=1,2$) and 
    $V_1(1)\calP V_1(-1)=V_2(1)\calP V_2(-1)$. Suppose 
    that the generators of $V_1$ and $V_2$ are positive. Then $V_1=V_2$. 
\end{Lemma}

\begin{proof}
By Borchers theorem \cite{Bo1} we have $\Delta^{is}V_k(t)\Delta^{-is}
= V_k(e^{-2\pi s}t)$, $t,s\in\mathbb R$, where $\Delta$ is the 
modular operator associated with $(\calP,\Omega)$. We then have 
$\Ad V_1(t)(\calP) = \Ad V_2(t)(\calP) $, $t\geq 0$, because
\[
\Ad V_k(e^{-2\pi s})(\calP)
=\Ad \Delta^{is}V_k(1)\Delta^{-is}(\calP)
=\Ad \Delta^{is}V_k(1)(\calP)=\Ad \Delta^{is}(\calP_1),
\ s\in\mathbb R.
\]
Then $Z(t)\equiv V_2(-t)V_1(t)$, $t\geq 0$, is $\Omega$-fixing and 
implements an automorphism 
of $\calP$, thus commutes with $\Delta^{is}$. 
On the other hand $\Delta^{is}Z(t)\Delta^{-is}=Z(e^{-2\pi s}t)$, 
due to the above commutation relations, so $Z(t)=Z(e^{-2\pi s}t)$ 
for all $t\geq 0$ and all $s\in\mathbb R$. Letting $s\to\infty$ 
we conclude that $Z(t)= 1$, that is $V_1(t)=V_2(t)$, for $t\geq 0$ and 
thus for all $t\in\mathbb R$ because $V_k(-t)=V_k(t)^*$.
\end{proof}

The following table summarizes the basic structure in the above 
discussion.

\[
\begin
{tabular}{l||l|l|l||l|l|l||}
{\it space} &{\it orbit} & {\it flow}  & $\omega$ &{\it orbit }&{\it flow} 
& $\omega$\\
\hline
Minkowski & geodesic & translations & ground & hyperbola & boosts & KMS \\
\hline
de Sitter & geodesic & boosts & KMS & geodesic & $\mu$ & ground\\
\hline
\end{tabular}
\]

\section{Two-Dimensional de Sitter Spacetime}
\label{dS2}

\subsection{Geometric preliminaries}

Let us assume that $dS^{2}$ is oriented and time-oriented.  Following
Borchers \cite{Bo2}, a wedge $W$ at the origin (namely a wedge whose
edge contains the origin) in the Minkowski space $M^{d}$ is determined
by an ordered pair of linearly independent future-pointing lightlike
vectors $\ell_{1},\ell_{2}$; $W$ is the open cone spanned by
$\ell_{1},-\ell_{2}$ and vectors orthogonal to $\ell_{1},\ell_{2}$ (in
the Minkowski metric).  In order to make this correspondence $1:1$ one
can normalize the vectors in such a way that their time-component is
1.  Clearly such a pair determines and is determined by the $(d-2)$
oriented hyperplane which is orthogonal to $\ell_{1}$ and $\ell_{2}$
w.r.t. the Minkowski metric (the edge of the wedge).  In particular,
when $d=3$, it is determined by an oriented line $\zeta$ through the
origin, e.g. by requiring that $\ell_{1},-\ell_{2},v$ determine the
orientation in $M^{3}$ when $v$ is an oriented vector in $\zeta$. 
Denoting by $x$, $\tilde{x}$ the intersection points of $\zeta$ with
$dS^{2}$, with $x$ preceding $\tilde{x}$ according to the orientation,
it is clear that $\tilde{x}$ is the symmetric of $x$ w.r.t the origin,
therefore $x$ determines the wedge, so the map $x\mapsto W(x)$ is a
bijection between points of $dS^{2}$ and wedges.

Any point $x$ in $dS^{2}$ determines two lightlike lines given by the
intersection of $dS^{2}$ with the tangent plane at $x$ (the ruled
lines through $x$ of the hyperboloid).  Let us denote them by
$h_{r}(x)$, $h_{l}(x)$ in such a way that, if $v_{r},v_{l}$ are future
pointing vectors in $h_{r}(x)$, $h_{l}(x)$ respectively, the pair 
$(v_{r},v_{l})$ determines the given orientation of $dS^{2}$. 

Now let us consider an observer generating the wedge $W(x)$.  Then,
the sets $h_{r}(x)=h_{r}(x)\cup h_{r}(\tilde{x})$, $h_{l}(x) =
h_{l}(x) \cup h_{l}(\tilde{x})$ form a bifurcated Killing horizon for
$dS^{2}$ \cite{KW}, see also \cite{GLRV1}), the Killing flow being the
one-parameter group of pure Lorentz transformations associated to the
wedge $W(x)$.  and the set $\frH=h_{r}(x)\cup
h_{l}(\tilde{x})$ is the event horizon for $W(x)$, which splits in the
two components $\frH_{+}=h_{r}(x)$ $\frH_{-}=
h_{l}(\tilde{x})$.

Clearly any point $x\in dS^{2}$ determines a partition of the space
into 6 disjoint regions: $W(x)$ (the right of $x$), $W(\tilde{x})$
(the left of $x$), $V_{+}(x)$ (the closed future cone at $x$),
$V_{-}(x)$ (the closed past cone at $x$), $V_{+}(\tilde{x})$ (the
closed future cone at $\tilde{x}$), $V_{-}(\tilde{x})$ (the closed
past cone at $\tilde{x}$).

\bigskip

\begin{tabular}{l l}
    \epsfbox{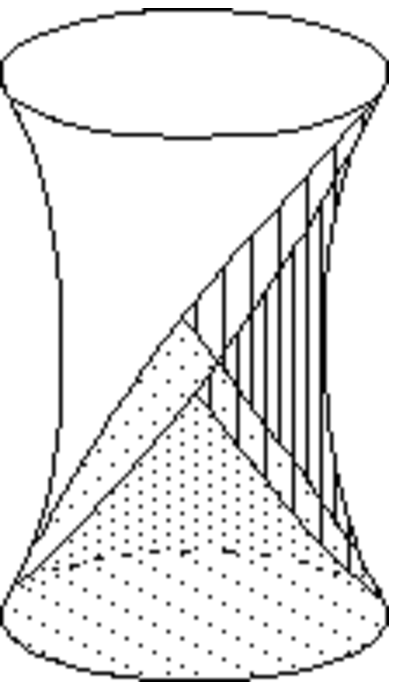} & \vbox{
    \hbox{\hsize=2.8in\hskip0.5cm \vbox{\lineskip=4pt\noindent Fig.~4. 
    Two-di\-men\-sion\-al de~Sitter space.  The whole marked area
    is the steady-state universe, whose boundary is the event horizon. 
    The striped area is the wedge region (static de Sitter spacetime), 
    whose boundary is the black-hole horizon.}} \hbox{\vbox{\vskip2.5cm}}}
\end{tabular}

\medskip

\begin{Lemma}\label{wedgeintersection}
	Two wedges $W(x)$, $W(y)$ have non-empty intersection if and
	only if $y$ belongs to $W(x)\cup W(\tilde{x})\cup V_{+}(x)\cup
	V_{-}(x)$.
\end{Lemma}

\begin{proof}
    The ``if'' part is obvious. Concerning the ``only if'' part, 
    assume that $y$ is in the future of $\tilde{x}$. Then $W(y)$ is 
    contained in the future of $W(\tilde{x})$. Since the latter is the 
    region in the future of the $h_{A}(x)$ horizon, while $W(x)$ is 
    contained in the past of $h_{A}(x)$, the thesis follows.
\end{proof}

Now we may characterize the sets that are intersections of wedges.

\begin{Lemma}\label{points}
	In $dS^{2}$, every non-empty open region $\calO$ given by an intersection of wedges, is
	indeed an intersection of two (canonically determined) wedges, or,
	equivalently, $\calO\in\tilde \calK$.
\end{Lemma}

\begin{proof} Let $\calO$ be an open region given by intersection of
wedges, and let $X(\calO)$ the set of points $x$ such that
$W(x)\supset\calO$.  Endow $X(\calO)$ with the partial order relation
of being ``to the right'', namely $x>y$ if $x\in W(y)$.  If $x,y\in
X(\calO)$ are not comparable, since $\calO\subset W(x)\cap W(y)$,
Lemma \ref{wedgeintersection} implies that one is in the future of the
other.  If $x$ is in the future of $y$, define $x\vee y$ as the
intersection of $h_{l}(x)$ with $h_{r}(y)$.  Clearly, if $\calO\subset
W(x)\cap W(y)$, then $\calO\subset W(x\vee y)$.  Therefore $X(\calO)$
is directed.  Since $\calO$ is open, the supremum of any ordered
subset in $X(\calO)$ belongs to $X(\calO)$, hence there exists a
maximal element.  Directedness implies that such maximal element is
indeed a maximum $L(\calO)$ (the leftmost point of the closure of
$\calO$).  Analogously we get a minimum $R(\calO)$ among the points
$y$ such that $W(\tilde{y})\supset\calO$ (the rightmost point of the
closure of $\calO$).  Clearly $\calO=W(L(\calO))\cap
W(\widetilde{R}(\calO))$.  Such a set is the double cone (possibly
degenerate, i.e. $\calO\in\widetilde\calK$) generated by the points
$F(\calO)=h_{r}(L(\calO))\cap h_{l}(R(\calO))$,
$P(\calO)=h_{l}(L(\calO))\cap h_{r}(R(\calO))$.
\end{proof}

We shall call $L(\calO),R(\calO)$ the spacelike endpoints of $\calO$, 
and $P(\calO),F(\calO)$ the timelike endpoints of $\calO$.

\subsection{Geometric holography}
\label{geoholo}
Now we fix the event horizon as the intersection of the plane 
$x_{0}=y_{0}$ with the de Sitter hyperboloid, the two components 
being $\frH_{\pm}=\{(t,t,\pm\rho):t\in\R\}$.
In the two-dimensional case, the orientation preserving isometry group 
of the de Sitter spacetime is isomorphic to $SO_0(2,1)$. On the other 
hand $SO_0(2,1)$ is isomorphic to $PSL(2,\mathbb R)$ and acts 
on (the one-point compactification of) $\frH_{+}$ or $\frH_{-}$. 
We shall construct holographies based on 
this equality.
 
The M\"obius group 
is the semidirect product of $PSL(2,\R)$ with 
$\Ze_{2}$. Let us chose the following generators for its Lie algebra 
$sl(2,\R)$:

\begin{equation}
    D=\frac12\left(\begin{matrix}1&0\\0&-1\end{matrix}\right),\quad
    T=\frac12\left(\begin{matrix}0&1\\0&0 \end{matrix}\right),\quad
    A=\frac12\left(\begin{matrix}0&0\\1&0 \end{matrix}\right).
\end{equation}

The following commutation relations hold:
\begin{equation}\label{commrel}
    [D,T]=T,\quad
    [D,A]=-A,\quad
    [T,A]=D.
\end{equation}

We consider also the following orientation reversing element of 
the M\"obius group:
\begin{equation}
    r=\left(\begin{matrix}-1&0\\0&1\end{matrix}\right).
\end{equation}

Let us observe that the following relations hold:

\begin{equation}\label{Rrel}
    rDr=D,\quad
    rTr=-T,\quad
    rAr=-A.
\end{equation}

We shall denote by $\b$ the usual action of the M\"obius group on
$\R\cup\{\infty\}$ as fractional linear transformations.  Then $\b(r)$
implements the reflection $x\mapsto -x$, $\b(\exp(tD))$ implements the
dilations, $\b(\exp(tT))$ implements the translations, and
$\b(\exp(tA))$ implements the anti-translations (see \cite{GL3}).

Now we consider the two immersions
\begin{align*}
    \psi_{\pm} :  \R &\to\frH_{\pm}\subset dS^{2}\\
     t &\mapsto(t,t,\pm\rho)
\end{align*}
of the real line in $ dS^{2}$ as $\pm$-horizon, and will look for 
actions $\a_{\pm}$ of the M\"obius group on $dS^2$
with the following property: 
whenever $\a_{\pm}(g)$ preserves $\frH_{\pm}$, then 
\begin{equation}\label{equivariance}
    \a_{\pm}(g)\psi_{\pm}(t)=\psi_{\pm}(\b(g)t).
\end{equation}
\begin{Lemma}\label{alpha}
    The previous requirement determines $\a_{\pm}$ uniquely, in 
    particular we have
    \begin{align*}
	\a_{+}(D)=\a_{-}(D)&=
	\left(\begin{matrix}0&1&0\\1&0&0\\0&0&0\end{matrix}\right)\\
	\a_{+}(T)=-\a_{-}(T)&=\frac12
	\left(\begin{matrix}0&0&1\\0&0&1\\1&-1&0\end{matrix}\right)\\
	\a_{+}(A)=-\a_{-}(A)&=\frac12
	\left(\begin{matrix}0&0&1\\0&0&-1\\1&1&0\end{matrix}\right)\\
	\a_{+}(r)=\a_{-}(r)&=
	\left(\begin{matrix}-1&0&0\\0&-1&0\\0&0&1\end{matrix}\right)
    \end{align*}
where $\a_{\pm}$ also denote the associated actions of $sl(2,\R)$. 
Moreover, the following relation holds:
\begin{equation}\label{Rrel2}
    \a_{-}(g)=\a_{+}(rgr).
\end{equation}

\end{Lemma}

\begin{proof}
    It is easy to see that the subgroup (globally) stabilizing
    $\frH_{+}$ coincides with the subgroup stabilizing $\frH_{-}$ and
    is generated by $\a_{+}(\exp(tD))$, $\a_{+}(\exp(tT))$, and
    $\a_{+}(r)$, as they are defined in the statement, therefore the
    identification is forced by eq.  (\ref{equivariance}) for these
    elements.  Eq.  (\ref{commrel}) implies then the formula for
    $\a_{+}(A)$.  The proof for $\a_{-}$ is analogous.  Relation
    (\ref{Rrel2}) immediately follows from the previous equations and
    relations (\ref{Rrel}).
    
\end{proof}

\begin{rem}\label{positivity}
    By Lemma \ref{alpha}, it follows that $\a_{+}(T)$ and $\a_{-}(T)$ 
    have opposite signs, thus, given a unitary representation $U$
    of $SO_0(2,1)$, the generator of $U(\a_{+}(\exp(tT))$ is positive 
    if and only if the generator of $U(\a_{-}(\exp(tT))$ is negative.
\end{rem}

Now we define two maps $\Phi_{\pm}$ from the set $\calW$ of wedges 
in $dS^{2}$ to the set $\calI$ of open intervals in (the one-point 
compactification of) $\R$ such that, for any element $g$ in the 
M\"obius group and any wedge $W$, one has
\begin{equation}\label{equivariance2}
    \Phi_{\pm}(\a_{\pm}(g)W)=\b(g)\Phi_{\pm}(W).
\end{equation}

\begin{Prop}\label{Phi}
    Let $W\in\calW$. Then $\partial 
    W\cap\frH_{+}\ne\emptyset\Leftrightarrow
    \partial W\cap\frH_{-}\ne\emptyset$. 
    The maps $\Phi_{\pm}$ are uniquely determined by the further 
    requirement that, for any such wedge, 
    \begin{equation}\label{boundarywedges}
	\Phi_{\pm}(W)=\psi_{\pm}^{-1}(\partial W).
    \end{equation}
    Moreover they satisfy
    \begin{align}
	\Phi_{\pm}(W')&=\Phi_{\pm}(W)'\label{eq:Phi1}\\
	\Phi_{+}(W)&=\b(r)\Phi_{-}(W'),\label{eq:Phi2}
    \end{align}
    where the prime ${}'$ denotes the spacelike complement in $dS^{2}$ and the 
    interior of the complement in $S^{1}$.
\end{Prop}

\begin{proof}
    Let us construct $\Phi_{+}$, the construction of $\Phi_{-}$ being
    analogous.  For notational simplicity we shall drop the subscript
    $_{+}$ in the rest of the proof.  Let $W_{0}$ be the wedge
    $W(0,1,0)$, according to the previous description.  Since the
    Lorentz group acts transitively on wedges, property
    (\ref{equivariance2}) may be equivalently asked for $W_{0}$ only. 
    Now eq.  (\ref{boundarywedges}) implies $\Phi(W_{0})=I_{0}$, where
    $I_{0}$ denotes the positive half line, hence we only have to test
    that equation $\Phi(\a(g)W_{0})=\b(g)I_{0}$ makes $\Phi$ well
    defined.  This is equivalent to show that if $\a(g)W_{0}=W_{0}$, 
    then $\b(g)I_{0}=I_{0}$. The stabilizer of $W_{0}$ is easily seen 
    to be generated by $\a(\exp(tD))$ and $\a(\hat{r})$, where 
    $\hat{r}=\left(\begin{matrix}0&1\\1&0\end{matrix}\right)$, since
    $$
    \a(\hat{r})=\left(\begin{matrix}-1&0&0\\0&1&0\\0&0&-1\end{matrix}\right).
    $$
    A direct computation shows that $\b(\exp(tD))$ and $\b(\hat{r})$
    stabilize $I_{0}$.  Now we show that eq.  (\ref{boundarywedges})
    is always satisfied.  Indeed, let $\partial
    W\cap\frH\ne\emptyset$, namely either $W=W(x)$ or
    $W=W(\tilde{x})$, with $x\in\frH$.  Then there exists $g$
    stabilizing $\frH$, either of the form $\exp(sT)$, or of the form
    $\exp(sT)r$, such that $W=\a(g)W_{0}$.  Eq.  (\ref{equivariance})
    then implies the thesis.
    
    Now we prove the (\ref{eq:Phi1}).  Indeed, by
    (\ref{equivariance2}), it is enough to prove it for only one
    wedge, e.g. $W_{0}$, where it follows immediately by
    (\ref{boundarywedges}). Concerning (\ref{eq:Phi2}), we have
    \begin{align*}
	\Phi_{+}(\a_{+}(g)W_{0})
	&=\b(g)\Phi_{+}(W_{0})=\b(g)\Phi_{-}(W_{0})\\
	&=\Phi_{-}(\a_{-}(g)W_{0})=
	\Phi_{-}(\a_{-}(r)\a_{+}(g)\a_{-}(r)W_{0})
	=\b(r)\Phi_{-}(\a_{+}(g)W_{0}').
	\end{align*}
\end{proof}

Let us observe that the above mentioned map trivially preserves inclusions, 
indeed no wedge is properly contained in another wedge of $dS^d$, 
while the inverse map does not.

Now we may pass to points. Indeed any point in $dS^{2}$ corresponds to 
a wedge: $x\mapsto W(x)$. Also, any interval in the 
one-point compactification of $\R$ determines its leftmost extreme: 
$I\mapsto \ell(I)$. Then we may define point maps as follows:
\begin{equation}\label{pointmap}
    \f_{\pm}(x)=\ell(\Phi_{\pm}(W(x))).
\end{equation}

\begin{Lemma}\label{Lemma:equivar}
    The point maps $\f_{\pm}$ are equivariant, namely
    \begin{equation}\label{equivariance3}
	\f_{\pm}(\a_{\pm}(g)x)=\b(g)\f_{\pm}(x).
    \end{equation}
\end{Lemma}

\begin{proof}
    Assume $g$ to be orientation preserving. Then 
    $\a_{\pm}(g)W(x)=W(\a_{\pm}(g)x)$ and $\ell(\b(I))=\b(\ell(I))$. 
    Therefore the result follows from (\ref{equivariance2}). Assume 
    now $g$ to be orientation reversing. Given $x\in dS^{2}$, we may 
    write $g$ as $h_{1}rh_{2}$, where 
    $\a_{+}(h_{2})x=x_{0}\equiv(0,0,\rho)$. Then eq. 
    (\ref{equivariance3}) reduces to 
    $\f_{+}(\a_{+}(r)x_{0})=\b(r)\f_{+}(x_{0})$, which is obvious. 
    The proof in the $-$ case is analogous.
\end{proof}

\begin{Thm} 
    The wedge maps $\Phi_{\pm}$ are induced by the point maps
    $\f_{\pm}$, namely $\Phi_{\pm}(W)=\{\f_{\pm}(x):x\in W\}$.  The
    point maps $\f_{\pm}$ are given by the holographic projections
    $$
    x\in dS^{2}\mapsto h_{\mp}(x)\cap\frH_{\pm},
    $$
where $\frH_{\pm}$ are identified with $\R$ as before.
\end{Thm}

\begin{proof}
    We prove the second statement first.  Indeed, it is sufficient to
    show that the preimage under $\f_{\pm}$ of a point $t$ in $\R$ is
    the ruled line $h_{\mp}(\psi_{\pm}(t))$.  Eq. 
    (\ref{equivariance3}) implies that this is simply the
    $\a_{\pm}$-orbit of the $\b$-stabilizer of $t$, and that we may 
    check the property for one point only, say $t=0$. The elements 
    of $SO(2,1)$ $\b$-stabilizing $0$ but not 
    $\a_{\pm}$-stabilizing $\psi_{\pm}(0)$ are of the form $\exp(sT)$, 
    and the orbit of $\a_{\pm}(\exp(sT))$ at $\psi_{\pm}(0)$ is 
    exactly $h_{\mp}(\psi_{\pm}(0))$.
    
    Now we prove the first statement in the $+$ case.  Let $x\in W$. 
    By equivariance, we can move $x$ and $W$ in such a way that
    $W=W(\psi_{+}t)$, and $x\in h_{-}(\psi_{+}(0))$. Then 
    the statement becomes $\f_{+}(x)\in\Phi_{+}(W)$, i.e. 
    $t<\f_{+}(x)$, but this is obvious since $x\in W$. The proof for 
    the $-$ case is analogous.
\end{proof}

The maps $\f_{\pm}$ may be considered as geometric holographies, 
namely projection maps from the de Sitter space to (some part of) the horizon 
preserving the causal structure and intertwining the symmetry group 
actions. Of course one can construct holography maps onto the 
conformal boundary as well, simply associating with any $x$ the 
intersection of $h_{\pm}(x)$ with the conformal boundary. 

\subsection{Pseudonets}\label{pseudo}
By a \emph{local conformal pseudonet} $\calB$ on 
a Hilbert space $\calH$ (or simply a local pseudonet) 
we shall mean here a map $\calB$ from the (proper, open, 
non-empty) intervals $\calI$ of $S^1$ to von Neumann algebras on 
$\calH$ with the following properties:
\begin{itemize}
\item \emph{M\"{o}bius covariance.} There exists a unitary 
representation $U$ of $PSL(2,\mathbb R)$ on $\calH$ such that
$U(g)\calB(I)U(g)^{-1}=\calB(gI)$, $g\in PSL(2,\mathbb R)$, $I\in\calI$.

\item  \emph{Vacuum with Reeh-Schlieder property.} There exists a 
unit, $U$-invariant vector $\Omega$, cyclic for each $\calB(I)$.
 
\item \emph{Interval KMS property.} $\Delta_{I}^{it}= U(\L_I(- 2\pi t))$, 
$I\in\calI$, where 
$\Delta_I$ is the modular operator associated with $(\calB(I),\Omega)$
and $\L_I$ is the one-parameter subgroup of $PSL(2,\mathbb R)$ of 
special conformal transformations associated with $I$, see \cite{BGL}.

\item \emph{Locality.} $\calB(I)$ and  $\calB(I')$ commute elementwise 
for every $I\in\calI$ (with $I'$ the interior of $S^1\setminus I$).
\end{itemize}

Note that we do \emph{not} assume {\emph{positivity  of the 
energy} (or negativity of the energy) nor \emph{isotony} (or 
anti-isotony).

Given a local pseudonet $\calB$ on the Hilbert space $\calH$, 
let $J$ be the canonical anti-unitary from $\calH$ to 
conjugate Hilbert space $\overline\calH$. 
We define the \emph{conjugate pseudonet} $\overline\calB$ 
on $\overline\calH$ by
\begin{equation*}
\overline{\calB}(I)=J\calB(I')J,\quad
\overline{U}(g) = JU(g)J,\quad
\overline{\Omega} = J\Omega \ .
\end{equation*}
We may define $\overline{\calB}$ directly on $\calH$ with the same 
vacuum vector by choosing a reflection $r$ on $S^1$ associated with 
any given interval $I_0$ (say $r: z\mapsto -z$) and putting
\[
\overline\calB(I)=\calB(rI')
\]
with the covariance unitary representation $\overline U$ given by
\[
\overline U(g) = U(rgr), \ g\in PSL(2,\mathbb R).
\]
In this case $\overline\calB$ depends on the
choice of $r$, but is well defined up to unitary equivalence.   
The second conjugate of $\calB$ is
equivalent to $\calB$.  $\calB$ is isotonic iff $\overline\calB$ is
anti-isotonic, and $\calB$ has positive energy iff $\overline\calB$
has negative energy.  Note that $\bar\calB$ is defined also if
$\Omega$ is not cyclic.

\begin{Thm}\label{pseudo-properties}
    Let $\calB$ be a local pseudonet.  
    \begin{itemize} 
	\item[$(i)$] Haag duality holds: $\calB(I)' = \calB(I')$,
	$I\in\calI$.
	\item [$(ii)$] If $\Omega$ is unique $U$-invariant, then each
	$\calB(I)$ is a type $III_1$ factor.
	\item [$(iii)$] $\calB$ is isotonic (resp.  anti-isotonic) iff
	it has positive energy (resp.  negative energy).
    \end{itemize} 
\end{Thm}
\begin{proof} 
    $(i)$ By locality, $\calB(I')$ is a von Neumann subalgebra of
    $\calB(I)'$, globally invariant with respect to the modular group
    Ad$\Delta_I^{-it}$ of $\calB(I)'$, hence $\calB(I')=\calB(I)'$ by
    Takesaki theorem due to the Reeh-Schlieder property of $\Omega$.
    
    $(ii)$ If $\Omega$ is unique $U$-invariant, then, as in
    \cite{GL3}, Ad$\Delta_I^{it}$ is ergodic on $\calB(I)$, and this
    entails the $III_1$-factor property.

    $(iii)$ If $\calB$ is isotonic, then positivity of the energy
    follows from the interval KMS property, see e.g. \cite{GLRV1}. 
    Conversely, if $U$ has positive energy, let us prove isotony. 
    Clearly it is enough to prove isotony for pairs $\tilde{I}\subset
    I$ having one extreme point in common, and, by $SO(2,1)$
    covariance, we need only one pair, say $I=(0,\infty)$,
    $\tilde{I}=(1,\infty)$, namely it is enough to show that
    translations $T(t)$ implement endomorphisms of $\calB(I)$ for
    positive $t$.  By a classical argument, positivity is equivalent
    to the positivity of the self-adjoint generator of the
    translations $T(t)$, therefore we have the four ingredients of the
    Borchers theorem: a vector $\Omega$, the vacuum, which is
    invariant for the representation, hence for the modular group
    $\Delta_{I}^{it}$ of $\calB(I)$ and for $U(T(t))$, the commutation
    relations between these one-parameter groups, the positivity of
    the generator of translations, and an (expected) implementation of
    $\calB(I)$-endomorphisms by $U(T(t))$ for positive $t$.  Davidson
    \cite{Davi} proved that the last property follows from the first
    three ones if the following holds: there exists an $\eps>0$ such
    that the vacuum is cyclic for the set $\calB(\eps)$ consisting of
    all the $x\in\calB(I)$ such that, for all $t\in(0,\eps)$, $U(T(t))
    x U(T(-t))$ is in $\calB(I)$ (Theorem 3 ibid.).  Now
    $\calB(\tilde{I})\subset\calB(1)$, hence the cyclicity follows.
    
    The equivalence between anti-isotony and negative energy is
    obtained by considering the conjugate pseudonet.
\end{proof}

Let us define the ``isotonized'' nets associated with $\calB$, resp. 
$\overline\calB$:
\[
\calB_{+}(I_0) = \bigcap_{I\supset I_0}\calB(I),\qquad
\calB_{-}(I_0) = \bigcap_{I\supset I_0}\overline\calB(I).
\]
Then $\calB_{\pm}$ is isotonic, thus it has positive energy (on the
vacuum cyclic subspace).  Moreover $\calB_{+}(I)$ is globally
invariant w.r.t. Ad$\Delta_{I}^{it}$ thus, by Takesaki theorem, there
is a vacuum preserving normal conditional expectation from $\calB(I)$
onto $\calB_{+}(I)$.  It is easy to check that
$$
\overline{\calB_{-}}(I_{0})=\bigcap_{I\subset I_{0}}\calB(I),
$$
hence $\overline{\calB_{-}}$ is expected in $\calB$ and
$\calB_{+}(I)\vee\overline{\calB_{-}}(I)\subset\calB(I)$.

\begin{Prop}\label{Bsplitting}
    If $\Omega$ is unique $U$-invariant, we have the von Neumann
    tensor product splitting
    $$
    \calB_{+}(I_1)\vee\calB_{-}(I_2) =
    \calB_{+}(I_1)\otimes\calB_{-}(I_2).
    $$
\end{Prop}
\begin{proof} 
    First we show that $\calB_{+}(I_1)$ and $\calB_{-}(I_2)$ commute
    for any $I_1, I_2\in\calI$.  As $\calB_{+}$ is a net, it is
    additive and we may assume that $rI_{1}\cup I_{2}$ has non-empty
    complement.  We may then enlarge $I_{2}$ in such a way that
    $rI_{1}\subset I_{2}$.  Then $\calB_{+}(I_{1}) \subset
    \calB(I_{1})$, and $\calB_{-}(I_{2}) \subset\overline\calB(I_{2}) =
    \calB(rI_{2})' \subset \calB(I_{1})'$, namely they commute. 
    Then, as in Theorem \ref{pseudo-properties} $(ii)$, the von
    Neumann algebras $\calB_{\pm}(I)$ are factors, hence they generate
    a von Neumann tensor product by Takesaki's theorem \cite{T}.
\end{proof}

\subsection{Holography and chirality}

Let $\calB$ a local pseudonet on $S^{1}$. Then we may associate with it a local 
net $\calA$ on the wedges of $dS^{2}$ as follows:
\begin{equation}\label{eq:holo}
    \calA(W)\equiv\calB(\Phi_{+}(W)).
\end{equation}
Clearly, given a pseudonet on $S^{1}$, we may obtain a net on the
double cones of $dS^{2}$ by intersection, and such net will satisfy
properties {\bf a)}, {\bf b)}, {\bf c)} on a suitable cyclic subspace. 
Conversely, given a net on $dS^{2}$, equation (\ref{eq:holo}) gives
rise to a pseudo net on $S^{1}$.

Let us consider the following property:
\begin{itemize}
    \item {\it Intersection cyclicity.} For any pair of intervals
    $I_{1}\subset I_{2}$, the vacuum vector is cyclic for the
    algebra
    \begin{equation}\label{cyclic}
	\calB(I_1,I_2)\equiv\bigcap_{I_{1}\subset I\subset I_{2}}\calB(I)
    \end{equation}
\end{itemize}

\begin{Thm} 
    The map (\ref{eq:holo}) gives rise to a natural bijective
    correspondence between:
    \begin{itemize}
	\item Haag dual nets $\calA$ on $dS^2$ (satisfying properties
	{\bf a)}, {\bf b)}, {\bf c)}, {\bf d)} in Sect. 
	\ref{QFTondS})
	\item Local pseudonets $\calB$ on $S^1$ satisfying
	intersection cyclicity.
    \end{itemize}
\end{Thm}

\begin{proof}
    We only have to check that, setting 
    $$
    \calA(\calO)=\bigcap_{W\supset\calO}\calA(W)
    $$
    for any double cone $\calO$, the intersection cyclicity is
    equivalent to the Reeh-Schlieder property for double cones.
    We shall show that
    \begin{equation}\label{coding}
	\calA(\calO)=\calB(\f_{+}(\calO),\b(r)\f_{-}(\calO)').
    \end{equation}
    Indeed, any double cone $\calO$ can be described as a Cartesian
    product: $\calO=I_{+}\times I_{-}$, where
    $I_{\pm}=\f_{\pm}(\calO)$.  Therefore, $W\supset\calO$ is
    equivalent to $\Phi_{\pm}(W)\supset I_{\pm}$.  Setting
    $I=\Phi_{+}(W)$ and making use of (\ref{eq:Phi2}), this is in turn
    equivalent to $I_{+}\subseteq I\subseteq \b(r)I_{-}'$.  In
    particular, since any double cone is contained in some wedge,
    $I_{+}, I_{-}$ give rise to a double cone $\calO=I_{+}\times
    I_{-}$ iff $I_{+}\subseteq \b(r)I_{-}'$.  The thesis follows.
\end{proof}

We showed that any Haag dual net on $dS^{2}$ can be holographically
reconstructed from a pseudonet on $S^{1}$.  Now we address the
question of when such a net is conformal.  Assuming intersection
cyclicity, let us denote by $\Delta_{I_{1}, I_{2}}$ the modular operator
associated with $(\calB(I_{1}, I_{2}),\Omega)$ for a pair of intervals
$I_{1}\subset I_{2}$.

\begin{Thm}
    Let $\calB$ be a local pseudonet on $S^{1}$ satisfying
    intersection cyclicity, $\calA$ the corresponding Haag dual net on
    $dS^{2}$.  Then $\calA$ is conformal if and only if, for any
    $I_{1}\subset L_{1}\subset L_{2}\subset I_{2}$,
    \begin{equation}\label{ci}
    \Delta_{I_{1}, I_{2}}^{it}
    \calB( L_{1}, L_{2})
    \Delta_{I_{1}, I_{2}}^{-it}
    = \calB(\L_{I_{1}}(- 2\pi t)( L_{1}),
    \L_{I_{2}}( 2\pi t)( L_{2}))
    \end{equation}
\end{Thm}

\begin{proof}
    Let us note that local modular covariance (for the inclusion
    $\tilde\calO\subset\calO$) can be rephrased, in view of equation
    \ref{coding}, as
    \begin{equation}\label{ci2}
    \Delta_{I_{1}, I_{2}}^{it}
    \calB( L_{1}, L_{2})
    \Delta_{I_{1}, I_{2}}^{-it}
    = \calB(\f_{+}(\L_{\calO}(t)\tilde\calO),
       \b(r)\f_{-}(\L_{\calO}(t)\tilde\calO)'),
    \end{equation}
    where $\calO$ and $\tilde\calO$ are determined by
    $\f_{+}(\calO)=I_{1}$, $\f_{-}(\calO)=\b(r)I_{2}'$,
    $\f_{+}(\tilde\calO)=L_{1}$, $\f_{-}(\tilde\calO)=\b(r)L_{2}'$.
    
    We want to show that for any double cone $\calO$, and any
    $x\in\calO$,
    $$
    \f_{+}(\Lambda_{\calO}(t)x) =
    \Lambda_{I}(t)\f_{+}(x)
    $$
    where $I=\f_{+}(\calO)$.  It is enough to show the property when
    $\partial\calO\cap\frH_{+}$ is non empty, since any other double
    cone can be reached via a transformation in the de Sitter group. 
    In this case, $I$ is identified with $\partial\calO \cap
    \frH_{+}$.
    
    First we observe that 
    $$
    \f_{+}(\Lambda_{\calO}(t)x) =
    \Lambda_{\calO}(t)\f_{+}(x),
    $$
    since, identifying de Sitter with Minkowski, $\Lambda_{\calO}$
    splits as the product of the action on the chiral components. 
    Since both are M\"obius transformations on $\frH_{+}$ leaving $I$
    globally invariant, they should coincide, possibly up a
    reparametrization. Finally, we find a conformal transformation 
    leaving $\frH_{+}$ globally stable and mapping $\calO$ onto a wedge 
    $W$, therefore it is enough to check the equality on a wedge, 
    where it follows by equivariance (\ref{equivariance3}). 
    
    As a consequence, whenever $\tilde\calO \subset \calO$, $\tilde I =
    \f_{+}(\tilde\calO)$, we get
    $$
    \f_{+}(\Lambda_{\calO}(t)\tilde\calO) =
    \Lambda_{I}(t)\tilde I.
    $$
    In an analogous way we get 
    $$
    \f_{-}(\Lambda_{\calO}(t)\tilde\calO) =
    \Lambda_{\f_{-}(\calO)}(t)\f_{-}(\tilde\calO).
    $$
    These equations show that relations (\ref{ci}) and (\ref{ci2}) are
    equivalent, therefore the thesis follows by Theorem
    \ref{equivalence of nets}.
\end{proof}

Now we study the geometric interpretation of the isotonized nets 
$\calB_{\pm}$.

We have seen that any double cone $\calO$ in $dS^{2}$ can be
represented as $\calO=I_{+}\times I_{-}$, where $I_{\pm} = \f_{\pm}
(\calO)$.  Then we may define the {\it horizon components} of a net
$\calA$ on $dS^{2}$ as the nets on $S^{1}$ given by
    \begin{equation}\label{chicomp}
	\calA_{+}(I)=\bigcap_{\calO:\f_{+}(\calO)\supset I}\calA(\calO)
	\quad,\quad
	\calA_{-}(I)=\bigcap_{\calO:\f_{-}(\calO)\supset I}\calA(\calO).
    \end{equation}

\begin{Thm}
    Let $\calA$ be a Haag dual net on $dS^{2}$, $\calB$ the
    corresponding pseudonet on $S^{1}$.  Then horizon components
    correspond to isotonized nets:
    \begin{equation}\label{components}
	\calA_{\pm}(I)=\calB_{\pm}(I).
    \end{equation}
    As a consequence the horizon components are conformal nets.
\end{Thm}

\begin{proof}
    Since $\calA$ is Haag dual, the chiral components may be 
    equivalently defined as
    \begin{equation}\label{chicomp2}
	\calA_{+}(I)=\bigcap_{W:\f_{+}(W)\supset I}\calA(W)
	\quad,\quad
	\calA_{-}(I)=\bigcap_{W:\f_{-}(W)\supset I}\calA(W),
    \end{equation}
    and the equality (\ref{components}) follows by eq. (\ref{eq:holo}).
\end{proof}

\begin{rem}
    We could have also defined the horizon restriction net for any
    component $\frH_{\pm}$ of the cosmological horizon, simply setting
    $\cap_{W\supset\psi_{\pm}(I)}\calA(W)$, for any interval
    $I\subset\R$.  In general it is a larger subnet than the horizon
    component $\calA_{\pm}$.
\end{rem}

Then we consider the conformal net on $dS^{2}$ given by
\begin{equation}\label{chinet}
    \calA_{\chi}(I_{+}\times I_{-})
    =\calA_{+}(I_{+})\vee\calA_{-}(I_{-}).
\end{equation}
\begin{Thm}\label{chir}
    $\calA_{\chi}$ is a conformal expected subnet of $\calA$,
    satisfying
    \begin{equation}\label{chiralsplit}
	\calA_{\chi}(I_{+}\times I_{-})
	=\calA_{+}(I_{+})\otimes\calA_{-}(I_{-}).
    \end{equation}
    Indeed it is the chiral subnet of the maximal conformal
    expected subnet of $\calA$.
\end{Thm}

\begin{proof}
    The tensor product splitting follows by Proposition
    \ref{Bsplitting}.  As $\calA_{\chi}$ is chiral conformal, it is
    immediate that it satisfies the local time-slice property, hence
    it is a Haag dual conformal net.  Thus it is expected by Prop. 
    \ref{dual}.
\end{proof}

We shall say that a net $\calA$ is chiral if it coincides with its 
chiral subnet.

The subnets $\calA_{\pm}$ may be considered as the chiral
components of $\calA$.  Indeed, they
correspond to the two chiral nets on the lightlike rays for a
conformal net on the two-dimensional Minkowski space.  
Therefore we shall say that $\calA$ is a \emph{chiral net} if
if it coincides with $\calA_{\chi}$.

The following table summarizes the chirality structure.
\[
\CD \fbox{Net on $dS^2$} @>\text{max.  conf.  subnet}> \text{Th.}\,
\ref{maximal conformal}>\fbox{Conformal net on $dS^2$} \\ @ V
\text{restriction} V \text{to horizons} V @ V \text{Th.}\,
\ref{equivalence of nets} V dS^d-M^d\ \text{conf.  equiv.} V \\
\fbox{Two conf.  nets on $\mathbb R$} @<\text{chiral components} <
\text{Th.}\, \ref{chir} < \fbox{Conformal net on $M^2$} \endCD
\]

We conclude this section characterizing the chiral nets on $dS^{2}$ with 
only one horizon component.

\begin{Thm}\label{hol}
    Let $\calA$ be a local net of von Neumann algebras on the de
    Sitter spacetime such that $H$ is positive, resp.  negative, where
    $H$ is the generator of the rotation subgroup.  Then the
    associated pseudonet $\calB$, resp $\overline\calB$ is indeed a
    local net, which holographically reconstructs $\calA$:
    $\calA(\calO)=\calB(\f_{\pm}(\calO))$.  In particular $\calA$ is
    conformal.
\end{Thm}

\begin{proof}
    If $H$ is positive, the pseudonet is isotonic, by
    \ref{pseudo-properties} $(iii)$.  Analogously, if $H$ is negative,
    the pseudonet is anti-isotonic, hence $\overline\calB$ is
    isotonic.  In both cases $\calA$ is chiral, hence conformal.
\end{proof}

\begin{Cor}\label{2d+E}
    The following are equivalent:
    \begin{itemize}
	\item The representation $U$ has positive (resp. 
	negative) energy
	\item $\calB_-(I)$ (resp.  $\calB_+(I)$) is trivial and
	$\calA_{\chi}(\calO)=\calA(\calO)$
	\item $\calA(I_{+}\times I_{-})=\calA_{+}(I_{+})
	\ ({\text{resp.}}\ =\calA_{-}(I_{-}))$
	\item $\calA$ is conformal and the translations on $\frH_{-}$ (resp. 
	on $\frH_{+}$) are trivial
    \end{itemize}
\end{Cor}
\begin{proof}
Immediate by the above discussion.
\end{proof}
We end up with a ``holographic'' dictionary:
\[
\boxed{\begin
{tabular}{ll}
$dS^2$ & $S^1$ \\
wedge $W$ & interval $I$ \\
double cone $\calO$ & pair of intervals $I_1\subset I_2$ \\
Haag dual nets $\calA$ on $dS^2$ & local pseudonets $\calB$ on 
$S^1$ \\
$SO(2,1)$ covariance & M\"{o}bius covariance \\
horizon (chiral) components $\calA_{\pm}$ &
isotonized nets $\calB_{\pm}$ \\
Reeh-Schlieder property for $\calO$ & intersection cyclicity \\
conformal invariance & property (\ref{ci}) \\
positive (negative) energy for $\calA$ & isotony (anti-isotony) for 
$\calB$\\
chirality & $\calB = \calB_+\otimes\overline{\calB_{-}}$ 

\end{tabular}}
\]

\section{Final Comments}
{\it Equivalence principle and dethermalization.} As is well known,
Einstein equivalence principle is a fundamental guiding principle in
General Relativity, although it is valid only at the infinitesimal
(i.e. local) level.  However, if one considers quantum effects, one
may notice a certain asymmetry, yet between inertial observers in
different spacetimes: the one in de Sitter spacetime feels the
Gibbons-Hawking temperature, while the one in Minkowski spacetime is
in a ground state.  One way to describe the dethermalization effect is
to say that it ``restores'' the symmetry: being a quantum effects, it
needs a quantum (i.e. noncommutative) description.  Only in the limit
case where QFT becomes conformal (a situation closer to general
covariance in classical general relativity) the dethermalization
effect is described by classical flows.  In the general case the
noncommutative geometry is encoded in the net of local algebras (that
takes the place of function algebras) and the dynamics is expressed in
terms of this net.

{\it Other spacetimes.} Although this paper has dealt essentially with
de Sitter spacetime, a good part of our description obviously holds in
more general spacetimes.  As mentioned, several spacetimes are
conformal to subregions of Einstein static universe.  For a
$d$-dimensional spacetime $\man$ in this class one can obviously
extend the analysis made in the $dS^d$ case: one can set up a
correspondence between local conformal nets on $\man$ and on $M^d$,
hence providing a KMS characterization of the conformal vacuum on
$\man$, and finding the evolutions corresponding to dethermalized
observers.  However, the partial geometric property of the
noncommutative flow with positive energy is established only in $dS^d$
case by using the large group of isometries of $dS^d$.

 In particular  we may consider a Robertson-Walker spacetime 
 $RW^d$. In the positive curvature case, 
 $RW^d$ is $\mathbb R\times S^{d-1}$ with metric 
 \[
 \text{d}s^2 = \text{d}t^2 - f(t)^2\text{d}\sigma^2 ,
 \]
 where $\text{d}\sigma^2$ is the metric on the unit sphere $S^{d-1}$
 and $f(t)>0$ (in the general case $S^{d-1}$ is a manifold of constant
 curvature $K=1,-1,0$.)  In this case we may also use the method of
 transplantation given in \cite{BMS}.
 
{\it Classification.} 
Recently \cite{KL}, diffeomorphism covariant local nets on the 
two-dimensional Minkowski spacetime, with central 
charge less than one, have been completely classified. 
By the conformal equivalence Th. \ref{equivalence of nets} 
one immediately translates this result on $dS^2$, 
namely one has a classification of the two-dimensional 
diffeomorphism covariant local nets on $dS^2$ with central charge less 
than one.

{\it Models, modular localization.} The methods in \cite{bglo3} 
provide a construction of (free) local nets on $dS^2$ associated with 
unitary representations of the de Sitter group $SO_0(d,1)$, and 
conformal nets on $S^1$ associated with unitary representations of 
$PSL(2.\mathbb R)$. The isomorphism between $SO_0(2,1)$ and $PSL(2,\mathbb 
R)$ gives the holography in these models and is at the basis of our 
general analysis.
\medskip

\noindent {\it Acknowledgements.} One of the authors (D.G.) wishes to 
thank the organizers and the participants to the E. Schr\"{o}dinger 
Institute program ``QFT on CST'', Vienna 2002, for the kind invitation 
and many helpful discussions.

\end{document}